# On the parameters affecting dual-target-function evaluation of single-particle selection from cryo-electron micrographs

Zhou Yu[a,b§], Wei Li Wang[a,c,d§], Luis R. Castillo-Menendez[a,c,d], Joseph Sodroski[a,c,d,e,f], Youdong Mao[a,c,d,g]*

[a]Intel® Parallel Computing Center for Structural Biology, Dana-Farber Cancer Institute, Boston, MA 02215. [b]Graduate School of Arts and Sciences, Department of Cellular and Molecular Biology, Harvard University, Cambridge, MA 02138. [c]Department of Cancer Immunology and AIDS, Dana-Farber Cancer Institute, Boston, MA 02215. [d]Department of Microbiology and Immunobiology, Harvard Medical School, Boston, MA 02115. [e]Department of Immunology and Infectious Diseases, Harvard School of Public Health, Boston, MA 02115. [f]Ragon Institute of MGH, MIT and Harvard, Cambridge, MA 02139. [g]School of Physics, Peking University, Beijing 100871, China.

[§]These authors contributed equally to this work.

*Corresponding author. Tel: +1 617 632 4358. Fax: +1 617 632 4338. E-mail address: youdong_mao@dfci.harvard.edu (Y. M.).

**ABSTRACT**

In the analysis of frozen hydrated biomolecules by single-particle cryo-electron microscopy, template-based particle picking by a target function called fast local correlation (FLC) allows a large number of particle images to be automatically picked from micrographs. A second, independent target function based on maximum likelihood (ML) can be used to align the images and verify the presence of signal in the picked particles. Although the paradigm of this dual-target-function (DTF) evaluation of single-particle selection has been practiced in recent years, it remains unclear how the performance of this DTF approach is affected by the signal-to-noise ratio of the images and by the choice of references for FLC and ML. Here we examine this problem through a systematic study of simulated data, followed by experimental substantiation. We quantitatively pinpoint the critical signal-to-noise ratio (SNR), at which the DTF approach starts losing its ability to select and verify particles from cryo-EM micrographs. A Gaussian model is shown to be as effective in picking particles as a single projection view of the imaged molecule in the tested cases. For both simulated micrographs and real cryo-EM data of the 173-kDa glucose isomerase complex, we found that the use of a Gaussian model to initialize the target functions suppressed the detrimental effect of reference bias in template-based particle selection. Given a sufficient signal-to-noise ratio in the images and the appropriate choice of references, the DTF approach can expedite the automated assembly of single-particle data sets.

Keywords: Automatic particle picking; Fast local correlation function; Cryo-EM; Maximum-likelihood estimator; Dual-target function validation; Single-particle reconstruction

# 1. Introduction

Image formation in electron microscopy is understood as the weak-phase approximation of thin, electron-penetrable objects [1]. The electron image formed after the objective lens is a convolution of the exit wave function passing through the object with the point spread function of the objective lens. The phase-contrast transfer function (CTF), which is the Fourier transform of the point spread function of the objective lens, gives rise to a tradeoff between the resolution transfer and the contrast transfer [2]. To image biomolecular structures in their native states by cryo-electron microscopy (cryo-EM), the molecules of interest are often flash frozen in a thin layer of amorphous ice. In cryo-electron micrographs, biomolecular objects are surrounded by imaging noise from electron scattering by the amorphous ice. Additional noise may be introduced in the process of electron signal transfer into the recording medium, such as shot noise in a CCD camera and electron counting noise in a direct electron detector. The weak phase approximation in image formation and the background ice noise often result in signal-to-noise ratios (SNR) of imaged biomolecules well below 1, commonly in the range of 0.01-0.3. Therefore, determination of cryo-EM structures of biomolecules at high resolution requires that numerous single-particle images, often on the scale of hundreds of thousands to a million, are acquired, aligned and averaged to remove background image noise in signal reconstruction.

Selection of single-particle images from noisy cryo-EM micrographs represents a major practical bottleneck in analyzing a large number of cryoEM images. Manual selection can be very time-consuming and is prone to errors resulting from subjective variables. Computerized particle selection has therefore been practically adopted for the assembly of a large number of single-particle images for cryo-EM structure refinement [3-5]. Over the past few decades, a number of computational tools have been developed toward the goal of automatic particle identification and verification [6-24]. For example, a template-matching approach has proven to be quite efficient in automated particle picking [16-18]. Recent automated particle selection approaches based on machine learning relieve the burden of post-picking manual selection [11,12,21]. Cross-correlation-based methods, such as a fast local correlation (FLC) algorithm, can successfully pick particles with low signal-to-noise ratios (SNRs) from cryo-EM micrographs [16-18]. However, the outcome of cross-correlation algorithms may be influenced by the alignment of noise to the template used as a reference, known as "reference bias" [25]. Recently, maximum likelihood (ML) optimization, which exhibited reduced susceptibility to reference bias compared to the cross-correlation algorithm [26,27], has been used to evaluate the homogeneity of the picked particles by repeating particle image alignment and unsupervised classification [28,29]. In principle, the use of two mathematically distinct target functions in signal recognition may serve as a test of the robustness of the image analysis and a verification of the detected signals, in that reference bias is not expected to be reproduced in the same way by two different target functions. The combination of one target function (FLC) for particle picking and the other target function (ML) for particle re-alignment has shown its potential in identifying the true signal in the selected images and, to a certain degree, suppressing the potential effect of reference bias (Fig. 1A) **[29-31].** Despite the application of this dual-target-function (DTF) evaluation paradigm in a number of single-particle cryo-EM structure determination tasks [29-31], it still remains uncertain how the DTF scheme performs at different SNR levels of the input data, how the choice of the particle-picking template affects the results of signal identification, and how the initialization of the second target function should be optimally implemented to suppress reference bias.

To address these gaps in knowledge, we evaluated how the performance of this dual-target-function (DTF) approach is affected by three variables: 1) the SNR of the cryo-EM data, 2) the template chosen for particle picking, and 3) the initialization reference used in ML alignment for signal verification. We quantitatively characterized the performance of the DTF approach on simulated micrographs exhibiting a wide range of SNRs. We performed comparative DTF studies with different references to investigate how the detrimental effect of reference bias incurred by the use of the first target function (FLC) may be suppressed by the application of the second target function (ML). Finally, we applied the DTF approach to real cryo-EM data of the 173-kDa glucose isomerase complex to substantiate experimentally that the use of a Gaussian template to initialize the target functions can lead to objective particle verification in semi-automated particle selection procedures.

## 2. Theory

### 2.1. Signal-to-noise ratio

The signal-to-noise ratio (SNR) compares the level of an observed signal to the level of background noise. The definition of the SNR varies in different contexts. For example, in electrical engineering, the SNR is defined as the ratio of signal power to the noise power, namely, $SNR = P_{Signal}/P_{Noise}$. Throughout this paper, the SNR is defined as the ratio of signal variance to the noise variance, conforming to convention in the field of cryo-EM image processing [2,32],

$$\text{SNR} = \sigma_{Signal}^2 / \sigma_{Noise}^2. \quad (1)$$

When the background noise has zero mean, its power $P_{\text{Noise}}$ equals its variance $\sigma_{Noise}^2$. In single-particle cryo-EM images, the expected signal, which contains information about the structure of the imaged molecule, may vary with spatial coordinates, making it a spatial variable instead of a constant. Therefore, when the signal mean is normalized to zero, $P_{\text{Signal}} = \sigma_{Signal}^2$ and the power ratio of signal to noise equals the variance ratio. Under these circumstances, the definition of SNR in equation (1) is equivalent to the electrical engineering definition. However, in the practice of cryo-EM data processing, the signal mean is not guaranteed to be zero, creating a distinction between the two definitions in most cases.

## 2.2. Target functions for signal alignment

Consider a set of $N$ images, each of which is a noisy, translated and rotated copy of the underlying 2D projection structure $A$. The $i$th image can be represented by

$$\boldsymbol{X}_i = R(\boldsymbol{\phi}_i)\boldsymbol{A} + \sigma \boldsymbol{G}_i, \quad i = 1, 2, \dots N, \quad (2)$$

where $X_i$ is the observed $i$th image of $J$ pixels each and with pixel values $X_{ij}$; $R(\boldsymbol{\phi}_i)$ denotes the in-plane transformation depending on the parameter vector $\boldsymbol{\phi}_i = (\alpha_i, x_i, y_i)$ that comprises a rotation $\alpha_i$ and two translations $x_i$ and $y_i$ along two orthogonal directions; $\boldsymbol{A}$ is the underlying signal with pixel values $A_j$ that is common to all images; $\boldsymbol{G}_i$ is the noise of a Gaussian distribution with a unity standard deviation, and further scaled by a scalar factor $\sigma$. Because the parameter vector $\boldsymbol{\phi}_i$ is experimentally unknown, the problem of image alignment is to determine the solution of a set of parameter vectors $\mathbf{\Phi} = \{\phi_i^{(n)}; i = 1, 2, \ldots N\}$ that allows an optimal estimate of the underlying true signal

$$A^{(n+1)} = \frac{1}{N}\sum_{i=1}^{N} R^{-1}(\phi_i^{(n)})X_i. \quad (3)$$

Here $R^{-1}(\phi_i^{(n)})$ is the reverse transformation that brings the image $X_i$ to the common orientation and position of $A$. This image alignment problem may be mathematically translated into different optimization problems. Two main types of mathematical translations have emerged in past studies [26,33]. In the first type, the image alignment problem was addressed by maximizing the squared magnitude of the summed images [33],

$$L(\mathbf{X}, \mathbf{\Phi}) = \left|\sum_{i=1}^{N} R^{-1}(\phi_i)X_i\right|^2. \quad (4)$$

The maximum of this function is equivalent to the minimization of the least-squares target:

$$L'(\mathbf{X}, \mathbf{\Phi}) = \sum_{i=1}^{N} \|X_i - R(\phi_i)A\|^2. \quad (5)$$

A local minimization of this function can be obtained by iteratively maximizing the cross-correlation between each image and the average,

$$\phi_i^{(n+1)} = \arg\max_\phi [X_i \cdot R(\phi_i)A^{(n)}], \quad i = 1, 2, \ldots N. \quad (6)$$

Here the dot indicates an inner product between two images $X \cdot A = \sum_{k=1}^{J} x_k a_k$. An approximate solution may be obtained by iteratively estimating the underlying signal $A^{(n)}$ and the alignment parameter $\phi_i^{(n)}$ according to equations (4) and (6).

In the second type, the image alignment problem was interpreted as a maximum-likelihood estimate (MLE) of the signal $A$, that is, the maximization of the probability function

$$\mathcal{L}(\Theta) = \prod_{i=1}^{N} P(X_i|\Theta). \quad (7)$$

Here $P(X_i|\Theta)$ is the probability density function of observing the image $X_i$ given the set of model parameters $\Theta = (A, \sigma, \xi)$, where $\xi$ characterizes the statistics of $R(\phi_i)$. In this case, the alignment parameters $\Phi = \{\phi_i; i = 1, 2, \ldots N\}$ are treated as latent variables. The maximization of the probability function $\mathcal{L}(\Theta)$ is more conveniently replaced by its logarithm

$$L(\Theta) = \sum_{i=1}^{N} \ln P(X_i|\Theta) = \sum_{i=1}^{N} \ln \int P(X_i|\phi, \Theta) P(\phi|\Theta) d\phi. \qquad (8)$$

A local maximum of the log-likelihood function $L(\Theta)$ can be obtained by finding $\Theta$ at which the partial derivatives of $L(\Theta)$ are zero. The problem of finding the maximum likelihood can be numerically tackled through the expectation-maximization algorithm. This algorithm is an iterative method that alternates between an expectation (E) step, which computes the expectation of the log-likelihood evaluated using the current estimate for the model parameters, and a maximization (M) step, which computes model parameters maximizing the expected log-likelihood found in the E-step [26,27]. These estimates of parameters are then used to determine the distribution of the latent variables in the next E-step. In the E-step, given the observed data $X_i$ and the current estimates of model parameters $\Theta^{(n)}$, one calculates

$$Q(\Theta, \Theta^{(n)}) = E_{\Phi|X,\Theta}\left[\sum_{i=1}^{N} \ln P(X_i, \phi|\Theta)\right] = \sum_{i=1}^{N} \int P(\phi|X_i, \Theta^{(n)}) \ln\{P(X_i|\phi, \Theta) P(\phi|\Theta)\} d\phi. \qquad (9)$$

Under the assumption of a Gaussian distribution of the latent variables $\Phi = \{\phi_i; i = 1, 2, \ldots N\}$ and the observed signal, this gives rise to

$$Q(\Theta, \Theta^{(n)}) \propto \sum_{i=1}^{N} \int P(\phi|X_i, \Theta^{(n)}) \{-\frac{1}{2\sigma^2}\|X_i - R(\phi)A\|^2\} d\phi. \qquad (10)$$

In the M-step, one maximizes $Q(\Theta, \Theta^{(n)})$ with respect to the model parameters

$$\Theta^{(n+1)} = \arg\max_{\Theta} Q(\Theta, \Theta^{(n)}), \qquad (11)$$

which corresponds to minimization of a weighted least-squares target with weight $P(\phi|X_i, \Theta^{(n)})$ for each image. Note that this is in marked contrast to equation (5). The estimate of signal is then a weighted average including contributions from all possible $\phi$ for every image $X_i$

$$A^{(n+1)} = \frac{1}{N}\sum_{i=1}^{N} \int P(\phi|X_i, \Theta^{(n)}) R^{-1}(\phi) X_i d\phi. \quad (12)$$

All other model parameters in $\Theta^{(n+1)}$ are updated in the M-step as similarly probability-weighted averages (Sigworth 1998).

The mathematical relationship and differences between the two types of image alignment approaches are considered. First, in recovering the signal $A$, the latter approach uses a probability-weighted average instead of the deterministic average used in the former approach, as illustrated by the differences between equations (3) and (12). Second, if one assumes that the estimate of hidden variables $\Phi$ is deterministic instead of probabilistic, $P(\phi_i|X_i, \Theta^{(n)})$ adopts the form of a Dirac δ-function. Under this condition, the maximization of the log-likelihood function shown in equation (10) is simplified to the minimization of the least-squares target shown in expression (5), instead of the probability-weighted least-squares target in equation (10); at the same time, the estimate of the signal by equation (12) would reduce to equation (3). Third, despite this conditional equivalence in terms of numerical optimization, the two approaches adopt essentially different target functions that include different variables and parameters, as is evident by comparison of equations (6) and (9). Importantly, all model parameters $\Theta = (A, \sigma, \xi)$ are re-estimated during each iteration of optimization in the latter approach, whereas only one type of model parameter, $A$, is re-estimated during the course of optimization in the former approach.

Previously proposed solutions to the particle-picking problem were mostly derived from the first type of image alignment approach. In a typical case, the locally normalized correlation function is calculated between a search object $S$ (template) and target micrograph $T$ under the footprint of a mask $M$ [16]:

$$C_L(x) = \frac{1}{P}\sum_{i=1}^{J} \frac{(S_i-\bar{S})M_i(T_{i+x}-\bar{T})}{\sigma_S \sigma_{MT}(x)}, \qquad (13)$$

where $\bar{S}$ and $\sigma_S$ are the average and standard deviation of the search object $S_i$; $\bar{T}$ and $\sigma_{MT}$ are the local average and standard deviation of $T$ within the footprint of mask M; $x$ is the position of the footprint of mask M, and P is the total number of non-zero points inside the mask. If $\bar{S}$ and $\sigma_S$ are set to zero and unity, respectively, equation (12) reduces to

$$C_L(x) = \frac{1}{P\sigma_{MT}(x)}\sum_{i=1}^{J} S_i M_i T_{i+x}. \qquad (14)$$

The local standard deviation of $T$ can be calculated by

$$\sigma_{MT}^2(x) = \frac{1}{P}\sum_{i=1}^{J} M_i T_{i-x}^2 - [\frac{1}{P}\sum_{i=1}^{J} M_i T_{i-x}]^2. \qquad (15)$$

This and other similar implementations of a particle-picking strategy have been collectively referred to as "template matching". As the image size of $S$ is much smaller than that of $T$, the local cross-correlation is calculated with the mask $M$ raster-scanning across the whole micrograph to produce a cross-correlation map. The local maximum in the correlation map is identified, ranked and used to indicate the position of the picked candidate particle image. The fast local correlation (FLC) function expressed in equation (14) has led to a more efficient implementation of a computational particle-picking procedure [16-18]. As explained above, the FLC target function notably differs from the MLE function in signal recognition; the two functions numerically respond distinctly to noise [26].

**2.3. Over-fitting and reference bias**

As noise can self-correlate to create a false-positive fit to a target function, over-fitting of noise can potentially afflict any target function or computational algorithm. This can be a barrier for the detection of weak signals in the midst of high background noise. In image analysis, when an experimental noisy image is compared with a reference image, the alignment parameters of the image (displacement and rotation) can be biased by the reference. This type of over-fitting of noise is generally referred to as reference bias or model bias. However, optimization of a multi-dimensional data set against different target functions can have dramatically different effects on over-fitting or reference bias. For example, the cross-correlation function exhibits a reference dependency that can persist in many iterations of optimization [25,26]. In contrast, the maximum-likelihood (ML) approach using a log-likelihood function regularly permits an escape from reference bias [26].

In image alignment, despite the aforementioned caveats, over-fitting can be avoided by the use of a featureless template, such as a Gaussian circle, or by employing a reference-free approach. On the other hand, if the reference used in image alignment does represent the intrinsic features of the signal present in the image, over-fitting is less likely to dominate, given a sufficient SNR. For a specific target function, it is important to define the lower bound of SNR beyond which the specific target function begins to fail in detecting or aligning signal.

**2.4. Rationale for the dual-target-function (DTF) approach**

As the SNR in cryo-EM images decreases, over-fitting and reference bias in a single target function can blur the "boundary" between signal and noise, creating a barrier for true signal to stand out. Nevertheless, it is mathematically prohibited that, under the same set of fitting parameters, the over-fitting of noise to one specific target function will necessarily be reproduced by another target function. Thus, the conceptual foundation of the DTF approach lies in an appropriate choice and use of a second target function that significantly differs from the first one; employing such a second target function should remove any potential over-fitting of noise resulting from the use of the first target function, allowing the true signal to be recovered. This DTF strategy can be used to detect and verify the signal present in cryo-EM micrographs.

Computerized procedures for signal detection in single-particle cryo-EM involve two steps: particle picking and particle verification. A number of algorithms have been developed to automate template-matching procedures for particle picking; these procedures require subsequent manual selection of particles, in some cases with the help of data clustering to expedite the rejection of false positives [22,23,34]. The majority of algorithms implementing template matching for particle-picking applications are based on the cross-correlation function, which calculates the normalized correlation between the template image and a local area of a micrograph. A disadvantage of the cross-correlation function is its sensitivity to noise, which can create false correlation peaks that do not result from real signal. However, these false, noise-based peaks of cross-correlation still retain the intrinsic statistical properties of noise; that is, their appearance in the 2D positions of a correlation map is random. When these pure noise images that are boxed out of a micrograph are aligned against a different target function, such as the ML estimator, the similarity of images indicated by the false correlation peak cannot be reproduced, due to the random nature of noise.

In the presence of signal and the absence of noise, the cross-correlation function and ML estimator both lead to the same solution for the image alignment problem (Sigworth 1998; Sigworth et al., 2010). However, in the presence of noise, the cross-correlation function demonstrates an increasing propensity to identify false-positive particles as the SNR decreases (Glaeser 2004; Zhu et al., 2004). In principle, although the ML estimator does not absolutely exclude the occurrence of false positives, its exhaustive probability search across parameter space substantially reduces the effect of false positives over the iterations of the expectation-maximization algorithm (Sigworth 1998). Therefore, following initial particle picking, particle verification by a reference-free ML alignment can be implemented (Figure 1); the generation of a clear 2D structure in the class averages, particularly if this structure is consistent with other available data, is strong evidence of the alignment of real signal in the images. When using reference-free alignment or using a featureless Gaussian circle as an initial reference, the imaging noise or false positives cannot dominate the ML optimization in the presence of sufficient signal. Therefore, an important question to be answered quantitatively in this study is, "What level of SNR is sufficient to permit the DTF approach to succeed?".

## 3. Methods

### 3.1. Practical implementation of the DTF approach

Throughout this study, the following implementation of the DTF approach was applied to 26 data sets of either pure noise or simulated low-contrast micrographs of the trimeric ectodomain of the influenza hemagglutinin (HA) glycoprotein [35], as well as an experimental data set of focal-pair micrographs of the 173-kDa glucose isomerase complex. An illustration of the DTF procedure is summarized in Fig. 1B.

Step 1: Particle picking by fast local cross-correlation. We used template matching by fast local cross-correlation implemented in SPIDER to pick particles [36]. The SPIDER script, lfc_pick.spi, has been studied in the case of the ribosome [18] and has served as a control for the recent development of a reference-free particle-picking approach [11]. This procedure applies the FLC function to particle recognition, following Roseman's (2003) approach (see section 2.2). In our study, we picked particles using single 2D templates, as described in the specific experiments below. Note that previous studies have shown that using the FLC function with a single template can pick many views of particles [18]. Nonetheless, it has been suggested that using more templates can potentially reduce the number of false positives that are picked [4,16-18].

Step 2: Candidate particle selection by the use of a threshold in the ranking of correlation peaks and manual rejection of obvious artifacts. The SPIDER particle-picking program (lfc_pick.spi) sorts and ranks the picked particles according to their correlation peaks, from high to low peak values. Upon sorting and ranking, the potential true particles often appear at higher correlation peak values and the pure noise images at lower correlation peaks. A threshold that approximately demarcates the boundary between the potential true particles and pure noise can be used to select the initial candidate particles, followed by manual inspection of each particle and rejection of obvious artifacts. The rejection of suspected artifacts and false positives can be done in a batch mode if the picked particles are clustered into groups (for example, by multivariate statistical analysis) [22,23,34].

Step 3: Particle validation by a reference-free ML alignment with single or multiple classes (Scheres et al., 2005; Scheres 2010). The ML-based approach for image alignment has been previously demonstrated to be quite resistant to reference bias after a sufficient number of iterations of optimization [26]. Image similarity measured by probability and subsequent class averages calculated by integration over all different probabilities are more sensitive to the presence of true signal [28]. Reference bias in particle selection would not be expected to persist through a number of iterations of multi-reference ML classification using a Gaussian circle as a starting reference. The particles belonging to the class averages that clearly exhibit the expected signal features are chosen for further processing; the particles in the class averages that are suspicious or apparently artifactual may then be discarded. This step provides an opportune checkpoint to efficiently remove non-particles in a batch mode. In the studies below, we specifically test the ability of ML alignment to extract signal from images with different SNR values and to suppress reference bias that was potentiated by template matching.

### 3.2. DTF testing of simulated and experimental noise micrographs

We first simulated 200 micrographs of only Gaussian noise by the SPIDER command MO (option R with Gaussian distribution). Each micrograph has dimensions of 4096 x 4096 pixels.

We then used one projection view of the ~11-Å human immunodeficiency virus (HIV-1) envelope glycoprotein trimer [37] as a template for particle picking from the simulated Gaussian-noise micrographs. The box size is 256 x 256 pixels. In each micrograph, about 20-25 boxed images of the highest local correlation peaks were selected to assemble a particle stack of 4485 images. After particle picking and selection, each particle image was scaled 4 times to 64 x 64 pixels (using xmipp_scale) and normalized (using xmipp_normalize) [38]. Subsequent ML alignment of a single class (using xmipp_ml_align2d) was repeated with three different starting references: (1) a noise image randomly chosen from the whole image stack; (2) a Gaussian circle, which follows a Gaussian distribution in radial intensity; and (3) an average of a random subset of the unaligned images that replicates the template used for particle picking.

To repeat the above DTF test on real experimental ice noise, we imaged a cryo-grid that was composed only of buffer solution and contained no protein sample. The composition of the buffer was 20 mM Tris-HCl, pH 7.4, 300 mM NaCl and 0.01% Cymal-6. This was the same buffer used for maintaining the HIV-1 membrane envelope glycoprotein trimer in solution during the cryo-EM data collection for its structural analysis [37,39]. The cryo-grid was made from a C-flat holey carbon grid by FEI Vitrobot Mark IV (FEI, OR, USA). The data were collected on an FEI Tecnai G2 F20 microscope operating at 120 kV, with a Gatan Ultrascan 4096 x 4096 pixel CCD camera, at a nominal magnification of 80,000. We selected 218 micrographs of pure ice noise collected in one cryo-EM session. The same particle-picking procedure performed with the simulated Gaussian noise micrographs (see above) was applied to the experimental ice noise micrographs, with the same HIV-1 envelope glycoprotein trimer template. After particle picking, the apparent ice-crystal contaminants were manually rejected from the particle set, leaving only images from amorphous ice noise. By selecting only about 10-25 boxed images of the highest local correlation peaks from each micrograph, a particle stack of 4591 images was assembled and was subjected to the same ML alignment as described above for the data from the simulated Gaussian noise micrographs. These DTF tests on both the simulated and experimental pure noise micrographs (Fig. 2) serve as controls for the subsequent examination of the effect of SNR on the success rate of the DTF approach.

### 3.3. DTF testing of simulated micrographs

We simulated 120 micrographs of noiseless particles corresponding to the crystal structure of the influenza A virus hemagglutinin (HA) glycoprotein ectodomain (PDB ID: 3HMG) (using xmipp_phantom_create_micrograph) [35]. The simulation assumes a pixel size of 1.0 Angstrom and micrograph dimensions of 4096 x 4096 pixels. The contrast transfer function (CTF) was applied in the Fourier transform of the simulated noiseless micrographs with a separate SPIDER script. The CTF simulation assumes an acceleration voltage of 200 kV, a defocus of -1 μm, a spherical aberration Cs of 2.0 mm, an amplitude contrast ratio of 10% and a Gaussian envelope half width of 0.333 Å$^{-1}$. In each simulated micrograph, there are 323 HA molecules that assume random orientations. To add different levels of Gaussian noise to the noiseless micrographs, the standard deviation of the background of each micrograph was calculated and used as input to simulate a background Gaussian noise image that was added to the noiseless micrographs. This results in micrographs with Gaussian noise added to yield SNRs of 0.1, 0.05, 0.02, 0.01, 0.005, 0.002, 0.001 or 0.0005. A typical series of a simulated noiseless micrograph and the derived noisy micrographs at different SNRs is shown in Fig. 3 and Supplementary Fig. 1. The corresponding behaviors of the power spectra in Fourier space are compared in Fig. 4.

For the simulated micrographs at each SNR value, we conducted DTF tests using three different templates for particle picking, i.e., a Gaussian circle, one projection view of the influenza virus HA trimer filtered to 30 Angstroms, and one projection view of the HIV-1 envelope glycoprotein trimer filtered to 30 Angstroms (Fig. 5). Each set of micrographs with a given SNR and selected by a particular particle-picking template is treated as a separate case. Therefore, there are 8 x 3 = 24 cases studied and compared in our DTF tests. For each case, a stack of 38,760 particle images was assembled, based on a selection threshold of 323, from 120 simulated micrographs. The original box dimension for particle picking was 180 x 180 pixels. After particle picking and selection, each particle image was first scaled 3 times to a dimension of 60 x 60 pixels and normalized for the background noise, then subjected to multi-reference ML classification into 5 classes, using two different initial references: (1) the unaligned average of a randomly selected subset of particles; and (2) a Gaussian circle, which follows a Gaussian distribution in radial intensity.

**3.4 DTF tests on experimental cryo-EM data**

We collected a real cryo-EM data set of the glucose isomerase complex (Hampton Research, CA, USA), which has been used as a crystallization standard specimen. The molecular weight of the glucose isomerase complex (173 kDa) is less than that of the influenza virus HA trimer (224.6 kDa). Glucose isomerase therefore represents a good model to investigate the lower bound of molecular sizes suitable for DTF analyses. A 2.5-μl drop of 3 mg/ml glucose isomerase solution was applied to a glow-discharged C-flat grid (R 1.2/1.3, 400 Mesh, Protochips, CA, USA), plunged into liquid ethane and flash frozen using the FEI Vitrobot Mark IV. The cryo-grid was imaged in an FEI Tecnai Arctica microscope at a nominal magnification of 21,000 times and an acceleration voltage of 200 keV. We selected 95 focal pairs of micrographs collected with the Gatan K2 Summit direct detector camera (Gatan Inc., CA, USA), with a defocus difference of 1.5 μm and a pixel size of 1.74 Å. The actual defocus values of the micrographs were determined through CTFFind3 [40].

To perform DTF tests on this cryo-EM data, we assembled three particle stacks (22298, 20632 and 22828 particles) by using three different templates for particle picking, i.e., a Gaussian circle, one projection view of the glucose isomerase crystal structure (PDB ID: 1OAD) filtered to 30 Å, and one projection view of the HIV-1 envelope glycoprotein trimer filtered to 30 Å. Particle images of 90 x 90 pixels, picked by FLC, were phase-flipped to partially correct the CTF effect. The three stacks of particles were normalized for the background noise and subjected to multi-reference ML classification into 5 classes, using two different initial references: (1) the unaligned average of a randomly selected subset of particles; and (2) a Gaussian circle, which follows a Gaussian distribution in radial intensity.

## 4. Results

### 4.1. DTF tests on simulated and experimental noise

As a control experiment to investigate the ability of the DTF approach to resist reference bias, we conducted DTF tests on simulated micrographs that contain only Gaussian noise. A single 2D projection of the HIV-1 envelope glycoprotein trimer was used as a template for picking "particles" by FLC (Target Function A) (Fig. 2A). Images with the highest local correlation peaks were selected and subjected to ML alignment, using three different starting references for ML optimization (Target Function B). In the first DTF test, a raw pure noise image randomly chosen from the particle stack was used as a starting reference for ML optimization (Fig. 2B). Over more than 3000 iterations of ML alignment, no 2D structure resembling the particle-picking template was observed. The resulting average image in each iteration was still a random noise image. We then used a Gaussian circle as the starting reference to repeat the ML optimization (Fig. 2C). Again, the resulting average image contained only random noise but no observable 2D model. As a third starting reference for ML optimization, we used the average of template-selected particle images without any further alignment. Notably, this average closely resembles the HIV-1 envelope glycoprotein template used for particle picking (Fig. 2D), and apparently results from reference bias in template-based particle picking by the FLC target function. Using this average image as a starting reference for the ML alignment, the replica of the template fades out in the average image and nearly disappears upon the convergence of ML optimization. Thus, the DTF approach can remove reference bias associated with the alignment of pure noise during the particle-picking process, particularly when the ML verification is conducted using a random noise image or a Gaussian circle as a starting reference. Note that in the above-mentioned test, we have performed up to 3000 iterations of ML optimization. Such a prolonged optimization provides the computation with a greater opportunity to evade local optima and helps to establish the robustness of the convergence [26].

Next, we asked if the results observed with the simulated micrographs of Gaussian noise would be reproduced with images of actual cryo-EM noise resulting from amorphous ice. We repeated the aforementioned DTF tests on the data set assembled from experimental ice noise micrographs. When aligned by ML, starting with pure noise or Gaussian circle references, no structure was observed after more than 3000 iterations of optimization (Fig. 2, E and F). When the unaligned average of the template-selected images was used as a starting reference for ML alignment, the noise-biased template image faded, but was not completely removed, by optimization (Fig. 2G). Thus, images of experimental ice noise taken by a CCD camera reproduce the results seen for simulated Gaussian noise, supporting the notion that the experimental cryo-EM noise from amorphous ice basically exhibits Gaussian-like behavior [2]. Particle verification by ML with starting references of random noise or a Gaussian circle effectively removed reference bias arising from the alignment of simulated or experimental noise. Removal of reference bias was less effective when unaligned averages of the images picked by a specific structural template were used as starting references for the ML alignment.

### 4.2. Simulated micrographs with different SNRs

Next, we tested the FLC-based particle-picking program on a number of simulated micrograph sets. Different levels of Gaussian noise were added to the same simulated noiseless micrographs, each containing 323 particles of influenza virus HA trimers in random orientations, to create images with SNRs of 0.1, 0.05, 0.02, 0.01, 0.005, 0.002, 0.001 and 0.0005. Figure 3 shows a typical noiseless micrograph (Fig. 3A) and the micrographs with different SNRs derived from it (Fig. 3B-D, Supplementary Fig. 1). As expected, the visibility of particles is drastically diminished in the lower SNR ranges [41]. We applied a number of contrast enhancement techniques, including histogram normalization, contrast stretching, low-pass filtering and pixel binning to the simulated micrographs with different SNRs (Supplementary Fig. 2). We found that these approaches were insufficient to restore unambiguous visibility to particles when the SNR approaches 0.002 (Supplementary Fig. 2B). Because the loss of visibility creates difficulty in directly verifying the true and false positives in the same low-contrast micrograph in our particle-picking test, the original noiseless micrograph from which the low-contrast micrograph was derived was used to verify particle-picking performance (Supplementary Fig. 3).

Using the noisy micrographs containing the randomly oriented influenza virus HA trimers, we picked particles with three different templates (a Gaussian circle, one projection view of the influenza virus HA trimer, and one projection view of the HIV-1 envelope glycoprotein trimer). Figures 5A-C show the plots of the correlation peak versus the rank number of picked particles. Notably, when the Gaussian circle was used as a template (Fig. 5A), the plots corresponding to SNRs of 0.1, 0.05, 0.02 and 0.01 showed a clearcut drop-off in the value of the correlation peak at a rank of 323, the number of particles simulated in each micrograph [3]. These 323 picked particles with high correlation peak values were confirmed to be true positives. When the Gaussian circle was used to pick particles from micrographs with an SNR of 0.005, the plot of the correlation peaks still exhibited a discernible drop-off at N = 323, but with a much smoother edge (Fig. 5A). The drop-offs in correlation peak values were smoother and less prominent at lower SNR values (0.002, 0.001 and 0.0005). Using 323 as the threshold for particle selection, the number of false positives was less than 2% at an SNR of 0.005, and increased to approximately 7% at an SNR of 0.002 (Fig. 5D).

We evaluated the specificity of particle picking when using templates other than a Gaussian circle; i.e., one projection view of the influenza virus HA trimer itself and one projection view of the HIV-1 envelope glycoprotein trimer, which bears little similarity to the HA trimer (Fig. 5B and C). For both templates, clear drop-offs in the correlation peak-ranking plots at N = 323 were observed at SNR values of 0.005 and higher. Notably, in all cases of using different templates in the particle-picking test, the false-positive rate was below 2.5% at SNR values of 0.005 and above; there were no false positives at SNR values of 0.02 and greater (Fig. 5D). However, using the Gaussian circle template allowed better centering of picked particles than using the other two templates (Fig. 6 and Supplementary Fig. 3). Among the cases compared here, the centering of picked particles was the worst when a dissimilar 2D structure (the HIV-1 envelope glycoprotein trimer) was used as a template for micrographs with the lowest SNRs (0.005-0.0005) (Fig. 6 and data not shown). Apparently, particle recognition is less sensitive to the detailed shape of the particle-picking template than are the specificity and particle-centering accuracy. Thus, the use of a dissimilar template succeeded in particle recognition at large, but resulted in a greater mis-centering of the picked particles and more false positives at the lowest SNRs (0.005-0.0005).

### 4.3. DTF tests on the simulated low-SNR particle sets

We evaluated the ability of the DTF approach to verify the presence of signal in the particles selected from micrographs with different SNRs by different particle-picking templates. Using a threshold of 323 to select the particles with higher correlation peaks, we subjected the selected particles to multi-reference ML classification and averaging (Figs. 7 and 8). The particle sets selected from micrographs with different SNRs using different templates were treated and classified separately, and the results were compared among the different SNRs and different particle-picking templates. Strikingly, for those data sets derived from micrographs with SNRs of 0.002 and higher, after ML optimization, the class averages all recapitulated the projection views of the influenza virus HA trimer, no matter what type of particle-picking template was used. The ML optimization results using particles selected from micrographs with SNR values of at least 0.002 were comparable for those selected by the Gaussian circle template (Figs. 7A,D and 8A,D) and those selected by the dissimilar HIV-1 envelope glycoprotein trimer template (Figs. 7C,F and 8C,F). Evidently, the model used for the particle-picking template does not govern the outcomes of ML optimization when sufficient signal is present.

Of note, the DTF test intermittently succeeded in aligning true signal even at an SNR as low as 0.001. However, at low SNR values, the frequency of such successful alignments and the quality of the class averages produced dropped significantly, as expected. Thus, at the lowest SNRs (0.001 and 0.0005), the DTF procedure became inefficient in verifying signal for this data set of 38760 particles. Considering that an SNR of 0.001 is unusually low and often can be avoided experimentally, the DTF tests on the simulated low-contrast micrographs should be relevant to the analysis of real cryo-EM experimental data.

**4.4. DTF tests on real cryo-EM images of glucose isomerase**

To further substantiate the utility of the DTF approach, we applied DTF tests to a real cryo-EM data set of the173-kDa glucose isomerase complex. Focal pairs of micrographs were recorded on a Gatan K2 Summit direct detector camera in the electron counting mode. The first exposure was taken at a defocus between -1.0 and -3.0 μm. In this defocus range, the visibility of the 173-kDa complexes is marginal, posing difficulties for manual particle identification (Fig. 9). The second exposure was taken at a defocus between -3.0 and -5.0 μm. In this defocus range, the particles are more visible (Fig. 9). We then used FLC to pick particles directly from the micrographs of the first exposure, and used the second exposure to verify the particle selection from the first exposure. The templates used for particle picking were a Gaussian circle, one view of the glucose isomerase complex, and one view of the HIV-1 Env trimer. The three particle sets selected with different particle-picking templates were classified separately, using two different starting references, i.e., the unaligned average of randomly selected subsets, and a Gaussian circle. The DTF tests from all six cases successfully produced class averages that correspond to projection views of the glucose isomerase complex (Fig. 10 and Supplementary Fig. 4). Consistent with our observations in the above simulation studies, the use of a Gaussian circle as both the particle-picking template and the ML alignment reference performed as well or better than the other combinations in generating class averages corresponding to glucose isomerase projections (Fig. 10B). When the HIV-1 Env trimer was used as the particle-picking template and the unaligned average used as the starting reference for ML alignment, two class averages showed structures that were strongly biased by the particle-picking template (Fig. 10E); the other three class averages more closely reflected the low-resolution projection views of glucose isomerase, although some residual elements of the HIV-1 Env trimer persisted in the background. However, when the Gaussian circle was used as the starting reference of ML alignment, the particle-picking template of the HIV-1 Env trimer was no longer recapitulated in any of the converged class averages (Fig. 10F). Even when one of the class averages demonstrated indistinct features, perhaps indicating a clustering of non-particle false positives, the aligned average did not resemble the particle-picking template of the HIV-1 Env trimer (the second row of Fig. 10F). As discussed above, in the DTF scheme, these classes of particles can be discarded, providing an opportunity to cull non-particles in a batch mode. These results indicate that the DTF approach, when used with Gaussian references, can be successfully applied to experimental cryo-EM data of a 173-kD protein complex.

**4.5. Effect of reference bias in particle selection and its limitations**

The fitting parameters in the particle-picking problem are the X-Y coordinates of the particle box. The choice of template in particle picking appears to bias the coordinates of the boxes. As shown in Fig. 6, the selected particles were best centered when using the Gaussian circle as a template, whereas the particle boxes deviated most from the particle centers when the template was one projection view of the HIV-1 Env trimer, a template that does not reflect the intrinsic structures in the micrographs. Consequently, the average image of the picked particles after boxing and before alignment closely resembled the template image (See the columns with the starting references (S. Ref.) in Fig. 7). However, the template neither changes the true signal in the boxed particle images nor is used in signal alignment by the ML estimator, allowing objective signal validation by the second target function. When a random class average of particles picked by a Gaussian circle was used as a starting reference for ML optimization, the results (Fig. 7A, D, G and J) were comparable to those using the unaligned average of particles picked with the HA trimer as a starting reference (Fig. 7B, E, H and K). In these cases, upon convergence, the class averages either showed the projection views of the influenza virus HA trimer (if successful) or showed a blank noisy image (if failed) (See the columns showing the 100[th] iteration). The same behavior was observed on the real cryo-EM data set of the glucose isomerase complex (Fig. 10A and C).

The HIV-1 Env trimer particle-picking template differs considerably from the structures actually present in the tested images; thus, the ability of DTF tests to suppress reference bias can be evaluated by assessing the Fourier ring correlation (FRC) between the particle-picking template and the class averages as they evolve during the process of ML optimization. We performed this analysis on the pure ice noise data (Fig. 2F and G), the simulated data of the influenza virus HA trimer (Fig. 7C, F, I and L and Fig. 8C, F, I and L), and the real cryo-EM data set of the glucose isomerase complex (Fig. 10E and F), comparing the unaligned averages and the Gaussian circle as starting references for ML optimization (Fig. 11). FRC values above a 0.5 cutoff over a resolution range of 20-100 Å quantify the degree of reference bias. First, we analyzed the cases in which the HIV-1 Env trimer was used to pick particles and unaligned class averages were used as starting references for ML optimization (solid curves in Fig. 11). For all of these cases, the FRC curves show a significant correlation (> 0.5) in the low-resolution range (20-50 Å) at the beginning of the ML optimization (black solid curves in Fig. 11A-F). However, as ML optimization progresses, the FRC values decrease and the image of the particle-picking template diminishes (Figs. 7C, F, I, L and 10E). In the case of the simulated influenza virus HA trimer data at an SNR of 0.005, the frequency of FRC-0.5 drops to 0.02 Å$^{-1}$ upon convergence, indicating an efficient removal of reference bias (Fig. 11A). Correspondingly, the converged ML class averages efficiently recovered the projection views of the influenza virus HA trimer (Fig. 7C). At SNRs of 0.001 and lower (Fig. 7I and L), ML optimization failed to recover the projection views of the influenza virus HA trimer; at these low SNRs, noisy traces reminiscent of the particle-picking template remained in some of the converged class averages (Fig. 7I and L and Fig. 11C and D). These results are consistent with those obtained with images of pure noise (Figs. 2G and 11E), and suggest that reference bias from the particle-picking template might be more efficiently suppressed by performing the ML alignment starting with a Gaussian circle reference. Indeed, in all ML alignments performed with a Gaussian circle as a starting reference, the FRC curves show no significant correlation (>0.5) between the HIV-1 Env trimer template and the converged class averages at a resolution higher than 10 nm (dashed curves in Fig. 11). Consistent with these observations, as shown in Fig. 8C, F, I, L and Fig. 10F, the particle-picking template was not observed during the process of ML optimization. Thus, when a Gaussian circle was used as a starting reference for ML optimization, the converged class averages did not recapitulate the structure of the particle-picking template.

## 5. Discussion

### 5.1. Parameters affecting DTF performance

The DTF approach employs two independent mathematical functions, capitalizing on the sensitivity of FLC-based particle picking and on the resistance to reference bias of ML-based particle alignment. We systematically investigated the impact of the SNR of the micrographs and the choice of reference models on the performance of the DTF approach. As SNR decreases, the risks of reference bias and the introduction of noise into the structure increase. We tested the ability of the DTF approach to guard against these pitfalls. The control experiments with simulated micrographs of Gaussian noise demonstrated that the reference bias derived from the FLC function does not necessarily translate into reference bias in the ML function, in either reference-free alignment or when a Gaussian circle is used as a starting reference. This conclusion also applies to the alignment of experimental cryo-EM ice noise. Together, these control experiments laid the rational foundation for testing the ability of the DTF approach to detect and verify weak signals in low-contrast simulated and real micrographs.

The DTF tests presented in this study make a number of critical points. First, the FLC implementation in SPIDER successfully picks particles with SNRs as low as 0.002. At this SNR, manual particle picking is difficult. Together with previous studies [16-18], our results suggest that the FLC function is highly sensitive to the presence of very weak signal. Second, a Gaussian circle is as effective at picking particles as a single projection view of the imaged molecule. Third, the detrimental effect of reference bias resulting from FLC-based particle picking can be largely suppressed by ML-based alignment using a Gaussian circle as a starting reference. Thus, given sufficient SNR in the images, the combination of FLC and ML, implemented with Gaussian references, provides a highly sensitive, objective way to detect and verify signal in cryo-EM micrographs.

The SNR of the micrographs is a critical parameter influencing the success rate of the DTF approach. Fortunately, the recent development and application of direct electron detectors have dramatically improved the detective quantum efficiency of image recording and enabled computational correction of sample movement and drift [42-52]. The DTF approach will benefit from the increased SNR provided by these advances, and in turn may assist the application of new technologies to a wider range of proteins and imaging conditions.

**5.2. Differences between FLC and the projection-matching algorithm**

The requirements for template matching in the particle-picking process differ somewhat from those for projection matching in structure refinement. In projection matching, one needs to be able to detect the specific features that distinguish one projection view from another. The calculation of a cross-correlation in projection matching generally involves two images of similar dimensions. In the particle-picking problem, one aims to detect the general presence of particles regardless of the detailed structure of each particle. In FLC calculations, the local correlation may be between two images of different dimensions. Therefore, fast template matching in particle picking needs only to calculate a low-frequency correlation in Fourier space in a coarse-grained manner [16]. This property renders the performance of FLC-based particle picking relatively insensitive to changes in the specific shape of the template. Quantitative differences between the two approaches have been discussed previously [16]. In our study, we found that the use of a dissimilar structure as the particle-picking template only marginally increased the number of false positives. As a result, a Gaussian circle may be a preferred picking template in the initial stage of automated particle picking, thus avoiding any potential selection bias [4]. Once a data set has been vetted by DTF and other validation approaches, it should be feasible to use the initial reconstruction from the data set to repeat the particle picking with multiple templates that more closely resemble the structure in the data set [4, 22,23]. This re-iteration of particle picking and re-assembly of the particle data set can potentially recover a majority of the false negatives from the early phase of particle selection.

**5.3. False positives**

Although false-positive particles will inevitably be picked by the cross-correlation function, the percentage of false positives in the candidate particle pools can be reduced by manual curation on both an individual particle level and a class-average level [16-18,23,34]. A reference-free ML alignment that leads to a clear 2D structure in class averages should allow an unambiguous distinction between weak signal and strong noise. Under conditions of Gaussian reference-initiated ML alignment, the false positives from pure noise cannot dominate the image alignment. Instead, through unsupervised alignment by ML, it should be possible to restore the weak signal in the presence of a small fraction of false positives in the data set.

Removing all false positives will be unlikely in real experiments involving a very large data assembly in that the appropriate selection threshold is not known and may vary from micrograph to micrograph. If a drop-off is observed in the correlation-peak ranking plot, the threshold can be estimated from the ranking number where the drop-off occurs [3]. However, in real cryo-EM micrographs, there are often more or less ice contaminants or non-particle features, which may be picked and become false positives. These non-particle features often have stronger correlation peaks and are readily recognizable and can be manually rejected from the data set [18]. Alternatively, the non-particles may be rejected and discarded as whole classes after the ML classification, when the class averages clearly indicate the absence of expected structural features; for instance, the particle class whose average is shown in the second row of Fig. 10F can be discarded as a whole prior to further particle analysis.

**5.4. Caveats in the application of DTF to experimental data**

Our quantitative characterization of the capabilities of the DTF test studied ideal cases with synthetic data. Differences exist between simulated and real micrographs in both the particle and the noise components. Although our simulated particles exhibit heterogeneous 2D views due to random 3D projection, they adopt a homogeneous conformation, whereas real particles may exhibit heterogeneity in conformation, beam-induced movement, defocus values, local ice thickness and sample charging, among others. Our simulated low-contrast micrographs are free of ice contaminants, which are found to some extent in experimental cryo-EM micrographs. As the false positives derived from ice contaminants often have high correlation peaks, they can appear in the micrographs at a wide range of SNRs. Additionally, the background ice noise may also deviate from a strict Gaussian distribution. Thus, the application of the DTF approach to actual experimental cryo-EM micrographs may deviate from the simulated ideal behavior [3,18]. For example, the degree of the drop-off in the correlation peak-ranking plot may be less than ideal, or the level of DTF efficiency at different SNRs may be reduced by the above-mentioned heterogeneity in particles and/or background. Despite these hypothetical differences between real and ideal experiments, the mathematical principle behind the DTF approach remains true, i.e., the detrimental effect of over-fitting by the first target function (FLC) in particle picking can be suppressed by the second target function (ML) in signal alignment. The applicability of this principle to the analysis of real cryo-EM data is supported by our study of low-contrast images of the 173-kDa glucose isomerase complex (Figs. 9-11).

Several issues should be considered when applying the DTF approach to experimental data. First, ice contaminants are the most frequent false positives in FLC particle picking. Recent advances in applying machine learning to particle selection can largely remove these types of false positives, with little manual intervention [11] . Moreover, it is often straightforward to remove ice contaminants manually. Second, the selection threshold (N) representing the number of true-positive particles is not precisely known in real experiments. However, the experimental N can be approximately estimated from the protein densities in micrographs made in parallel under the same biochemical conditions and imaged using parameters that favor higher contrast, such as lower acceleration voltage, a smaller objective aperture and higher defocus. It is preferable to estimate the experimental N conservatively before applying it to lower-contrast micrographs. The availability of focal pairs of micrographs allowed an estimate of N for our DTF tests on the experimental data set of the glucose isomerase complex (Figs. 9 and 10). Third, experimental SNR is expected to fluctuate, in contrast to the fixed SNR used in our simulation studies. Therefore, image background normalization could increase the sensitivity in detecting weak signals. Fourth, in experimental micrographs, a subset of the particles are likely to be overlapping or in contact with one another, whereas the particles in our simulated micrographs were separated. Overlapping or touching particles picked from aggregates may be able to be removed after ML classification due to their different sizes or shapes. Although additional experimentation will be required to optimize DTF procedures for the analysis of real cryo-EM data, our study of the 173-kDa glucose isomerase complex indicates that such application is feasible.

Note that the SNR calculated for a whole micrograph is often lower than the SNR calculated from boxed single-particle images, given that there are more empty background areas in the micrograph than in appropriately boxed single-particle images. When extrapolating the results of this study to the SNR of single-particle images, the SNR of a whole micrograph should be multiplied by a factor of 2 to be equivalent to the SNR of boxed particle images.

Several caveats apply when extrapolating the results in this study to other cryo-EM data. First, although a Gaussian model works quite well for picking globular particles, it could be error-prone for particles with unique shapes and topology, such as ring-like and rod-like structures (Glaeser 2004). In this case, a model low-pass filtered at 60 Å, which follows the low-frequency features of particles, could be used as a particle-picking template. Second, our results do not completely negate the potential risk of using lower SNR data, nor do they encourage the use of critical SNR data. Rather, our study quantitatively defines the critical SNR where DTF, as currently applied, may succeed or fail for a given set of data, points out the optimal choices of parameters in DTF practice and provides a benchmark control and reference metric for diagnosing potential issues in analyzing noisy particle data by the DTF approach. In practice, optimization of sample quality and imaging conditions should always precede the optimization of parameters in single-particle data analysis. Third, the DTF approach does not bypass the requirement for manual verification and inspection of picked particles either at a single-particle or single-class level. Suspicious class averages after the Gaussian-initiated ML alignment and classification should be manually identified and removed selectively or as a whole class from the commissioned data set. Finally, the optimization of spectral SNR (SSNR) in single-particle images is not always compatible with the maximization of image contrast; the use of a larger defocus that increases low-frequency image contrast can introduce more zero crossovers in the resulting CTF that reduce the high-frequency SNR in the recorded particle images. It is important to use the SSNR as a major metric for image quality optimization, seeking a balance between improved SSNR and reduced contrast. Applying a higher imaging dose in the frame-packed video imaging mode of a direct electron detection camera can at least partly compensate for the low-frequency contrast loss associated with the use of lower defocus to improve the SSNR. Indeed, several recent high-resolution cryo-EM studies have used data with defocus values as low as -600 nm to -1.5 μm [53,54]. These caveats underscore the need for a better quantitative understanding of DTF performance in single-particle image analysis, justifying this initial study as well as future investigation of this subject.

## 6. Concluding remarks

In this work, we examined the effects of SNR and choice of references on the ability of the DTF approach to select and verify particles from noisy micrographs. We characterized the quantitative performance of FLC-based particle selection and ML-based particle verification over a wide range of SNRs. We quantitatively characterized the critical SNR where DTF performance begins to degrade. We found that the critical SNR is surprisingly small, as low as 0.002-0.005, given the size of the data set (38760 particles) tested in each case. The DTF approach, which combines the two target functions, represents a sensitive, objective way to assemble particles for downstream cryo-EM structure refinement. Importantly, reference bias from the FLC target function does not necessarily transfer to the ML target function, making possible the robust detection and objective validation of weak signal. In particular, we found that the use of a Gaussian model as a reference in the particle-picking and verification procedures incurs no apparent bias. The use of a Gaussian model is essentially as successful as using a projection of the known structure to extract true positives in the examined cases; therefore, a Gaussian circle is the preferred template for initial particle picking from low-contrast images. An ML-based classification in the verification step can then be used to establish whether the false positives overwhelm the true positives in the data set. When a non-Gaussian template is used for particle picking by the first target function, we found that the use of a Gaussian model to initialize the second target function, ML optimization, can largely suppress potential model bias from the particle-picking template. Finally, we demonstrated the applicability of the DTF approach to real cryo-EM data from the relatively small (173-kDa) glucose isomerase complex. The benchmarks learned in this study could be dependent on the target functions chosen. Therefore, caution should be exercised when other target functions not tested here are introduced as alternatives in the DTF scheme.

## Acknowledgements

The authors thank J. Jackson and T. Song for assistance in maintaining the high-performance computing system; C. Marks, A. Graham, A. Magyar and D. Bell for assistance in maintaining the imaging system; Y. McLaughlin and E. Carpelan for assistance in manuscript preparation. The experiments and data processing were performed in part at the Center for Nanoscale Systems at Harvard University, a member of the National Nanotechnology Infrastructure Network (NNIN), which is supported by the National Science Foundation under NSF award no.

ECS-0335765. The cryo-EM facility was supported by the NIH grant AI100645 Center for HIV/AIDS Vaccine Immunology and Immunogen Design (CHAVI-ID). This work was funded by an Intel academic grant (to Y.M.), by the National Institutes of Health (NIH) (AI93256, AI67854, AI100645 and AI24755), by an Innovation Award and a Fellowship Award from the Ragon Institute of MGH, MIT and Harvard (to Y.M.), and by gifts from Mr. and Mrs. Daniel J. Sullivan, Jr.

**Author Contributions**

Y.M. conceived the concepts, designed the experiments, and conducted the simulation study. W.L.W., L.R.C.-M., Y.M. conducted the cryo-EM experiments. Z.Y. conducted the DTF tests on the real cryo-EM data of the glucose isomerase complex. Y.M. and J.S. wrote the manuscript.

**Figure Legends**

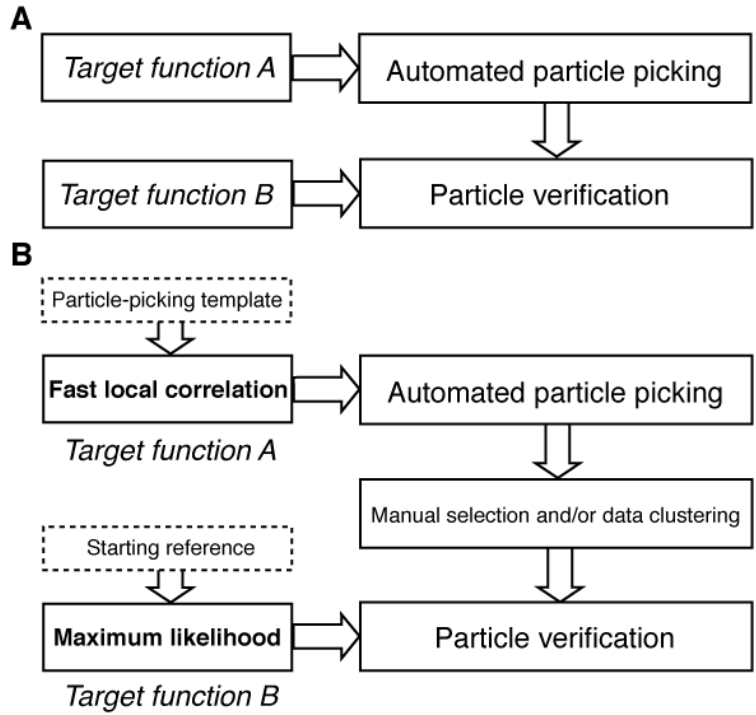

**Figure 1**. Strategy and implementation of the DTF approach. (A) The DTF approach involves the use of two different target functions. The first target function deals with particle detection and the second target function with particle verification. (B) The DTF approach used in this study combines fast local correlation (FLC) and maximum likelihood (ML) target functions, which are not mathematically equivalent or correlated. User-determined templates/references are shown in the dashed boxes, designated by the terms that will be used throughout this manuscript.

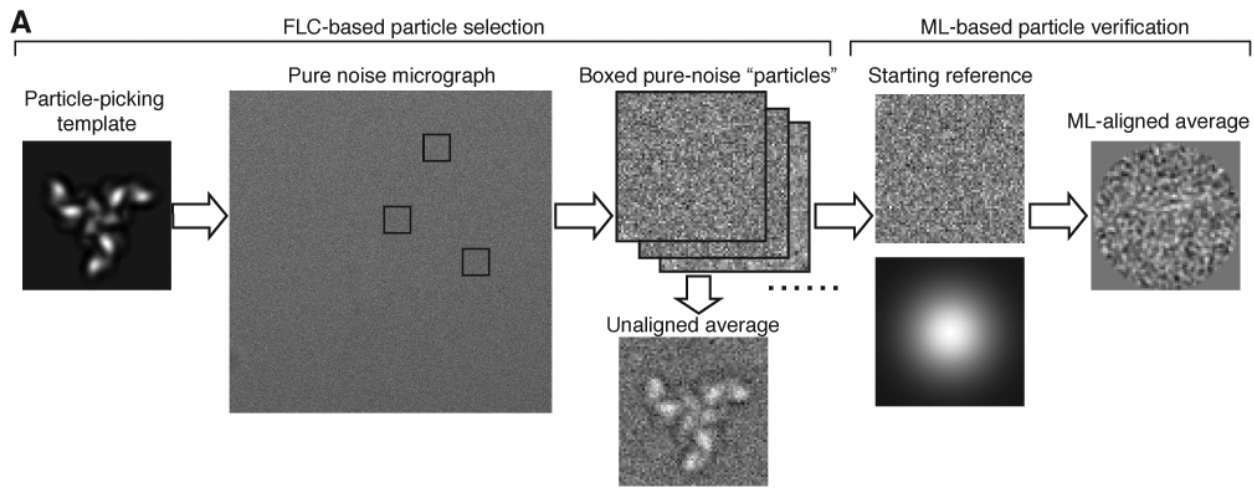

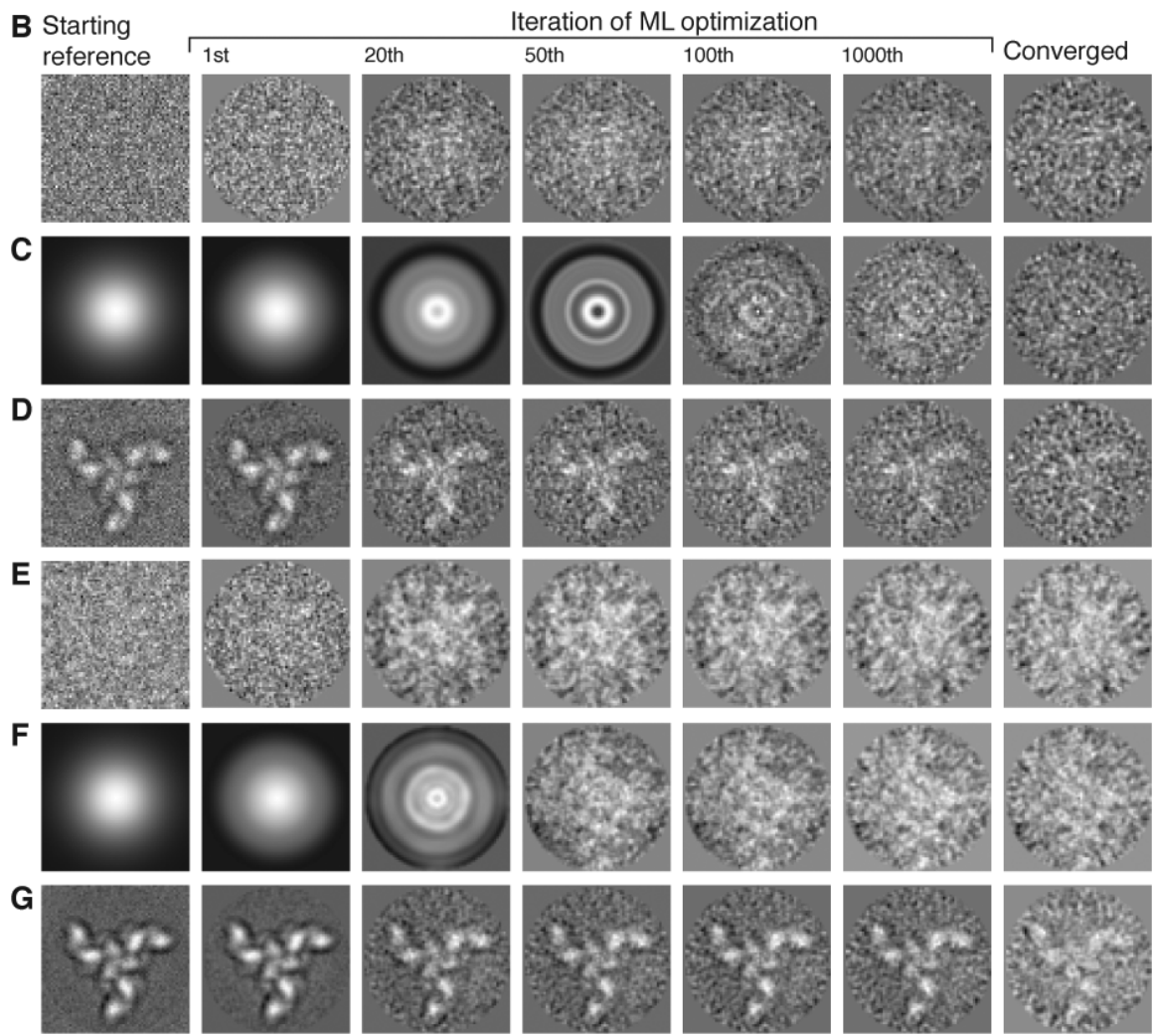

**Figure 2**. The DTF results for pure noise data, both simulated and experimental. (A) A schematic flow diagram shows that "particles" were picked by FLC from pure-noise

micrographs, using a single projection of the HIV-1 envelope glycoprotein trimer as a template. The picked particles were subjected to ML alignment, using different starting references. (B-D) The FLC-picked particle set, derived from the simulated Gaussian-noise micrographs, was aligned by ML, starting from a noise image randomly chosen from the particle set (B), a Gaussian circle (C), or the average of the picked particles without any further alignment (D). This starting reference for ML optimization is shown in the first column. Each row shows the history of the ML-aligned class averages at the indicated iterations of optimization, ending with the converged class average in the far right column. (E-G) The FLC-picked particle set, derived from the experimental ice noise micrographs, was aligned by ML, starting from a noise image randomly chosen from the particle set (E), a Gaussian circle (F), or the average of the picked particles without any further alignment (G). The averages shown in (D) and (G) appear as an FLC-generated replicate of the 2D template used in the particle picking.

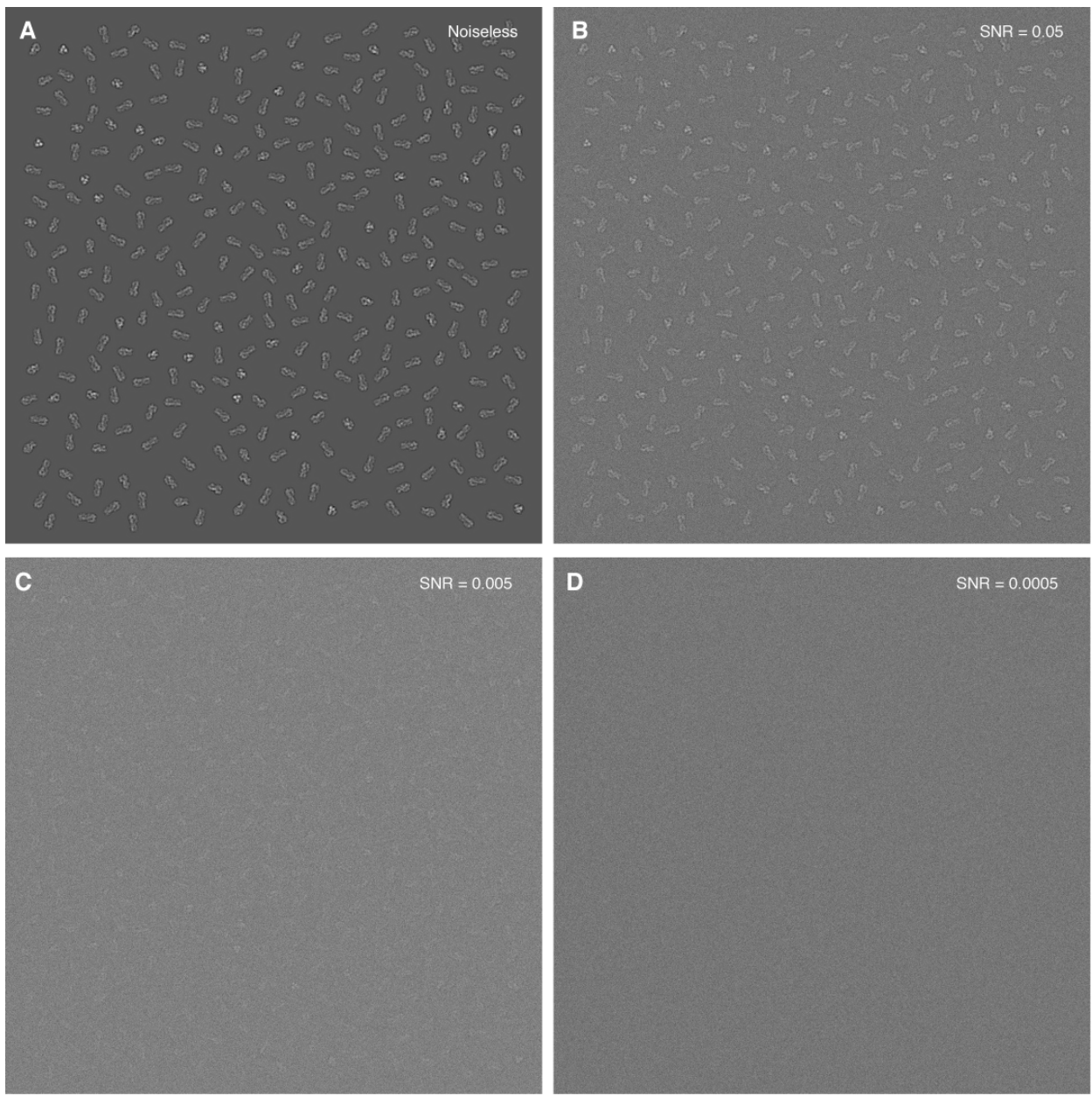

**Figure 3**. The simulated micrographs with different SNRs. (A) An example is shown of a simulated noiseless micrograph containing projection views of the influenza virus HA trimers in random orientations. (B-D) A different level of Gaussian noise was added to the noiseless micrograph shown in (A) to simulate noisy micrographs at an SNR of 0.05 (B), 0.005 (C), and 0.0005 (D).

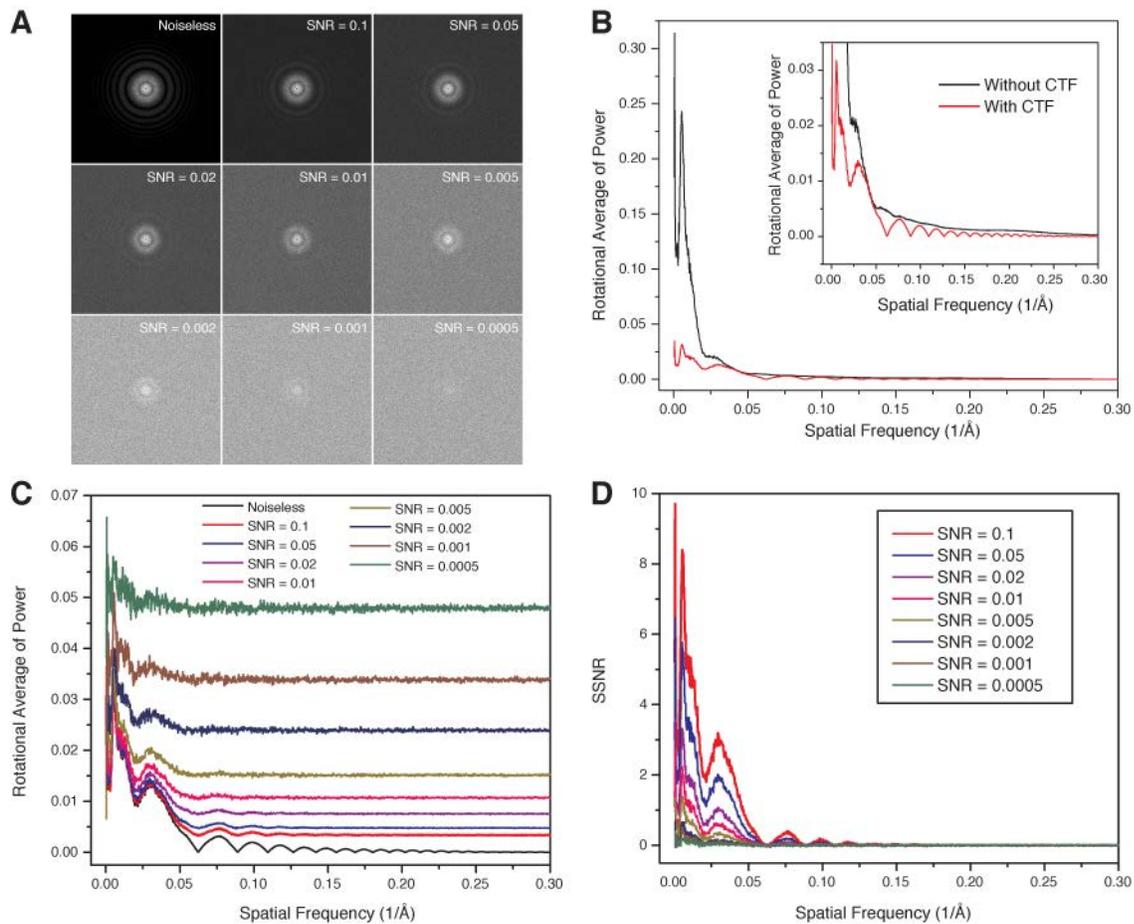

**Figure 4**. The Fourier behavior of the simulated micrographs. (A) The power spectra of the simulated micrographs with different SNRs. (B) The rotational average of the power spectrum of the noiseless micrograph before and after applying the CTF effect. (C) The rotational average of the power spectra of the simulated noisy micrographs. (D) The spectral signal-to-noise ratio (SSNR) of the simulated noisy micrographs.

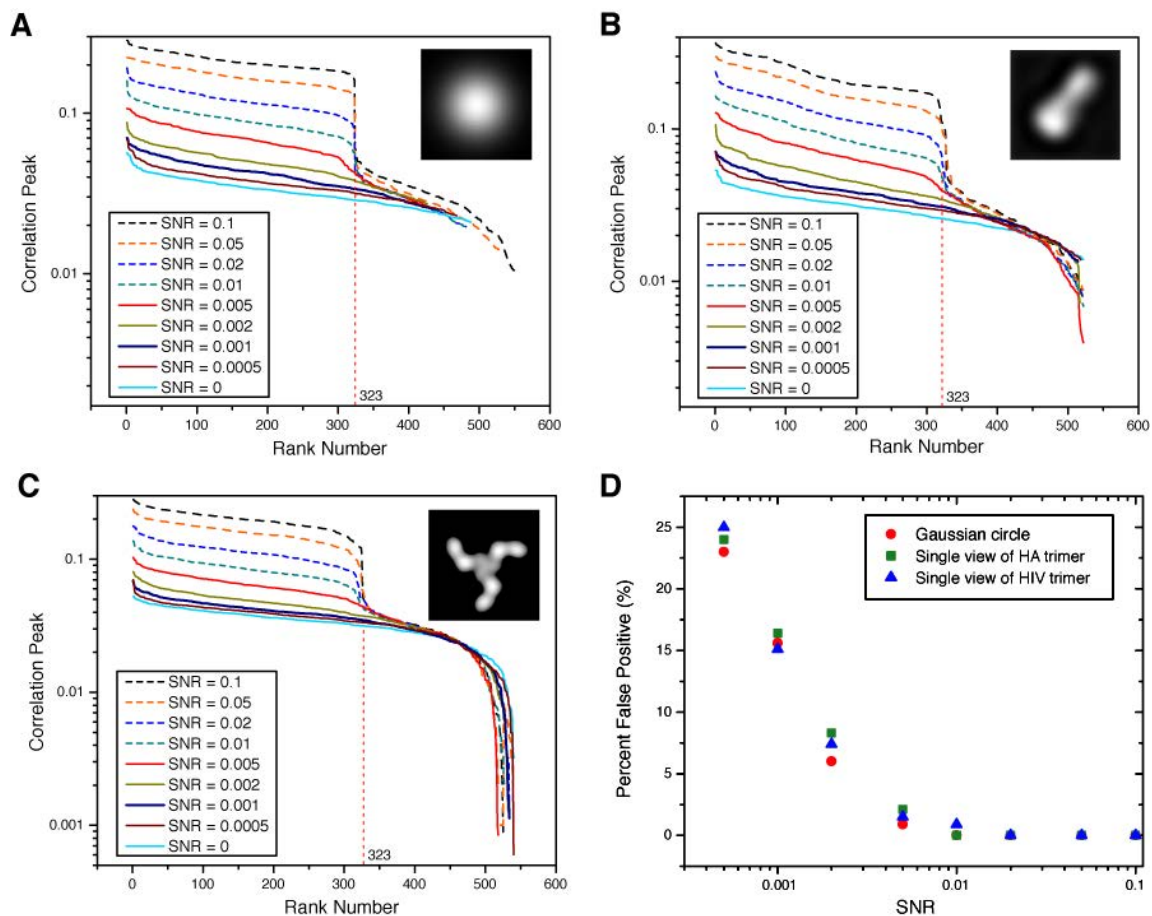

**Figure 5**. The correlation peak-ranking plots and differentiation of true-positive and false-positive particles in FLC-based automated particle picking. The correlation peak-ranking plots corresponding to different SNRs, using three different particle-picking templates: (A) a Gaussian circle, (B) one projection view of the influenza virus HA trimer, and (C) one projection view of the HIV-1 envelope glycoprotein trimer. The particle-picking templates are shown in the insets. All plots are from the noisy particle micrographs derived from the same simulated noiseless micrograph of the influenza virus HA trimer. Note that the position of the drop-off in correlation peak values corresponds to 323, the number of actual influenza virus HA trimers in the simulated micrographs. (D) Rate of false positivity in particle picking. The plots of false positives in particle picking by the three different templates are shown, indicating that the specificity of FLC

particle picking is highly dependent on the SNR, and is also affected to a lesser extent by the choice of the 2D template. Below a critical SNR range (0.002-0.005), the percentage of false positives rises considerably.

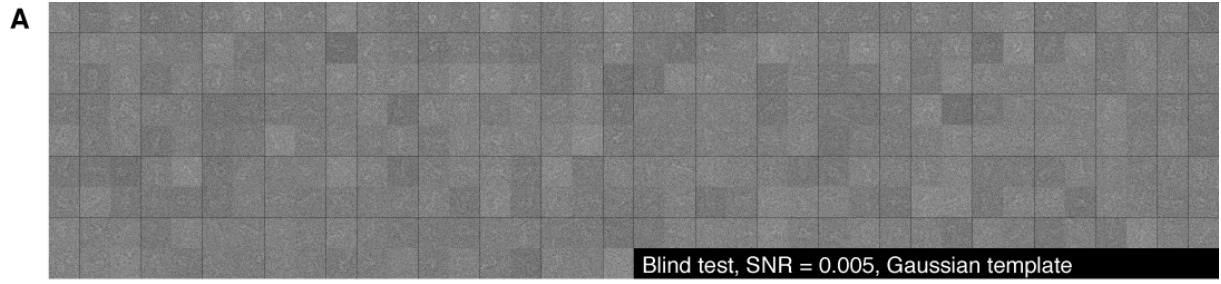
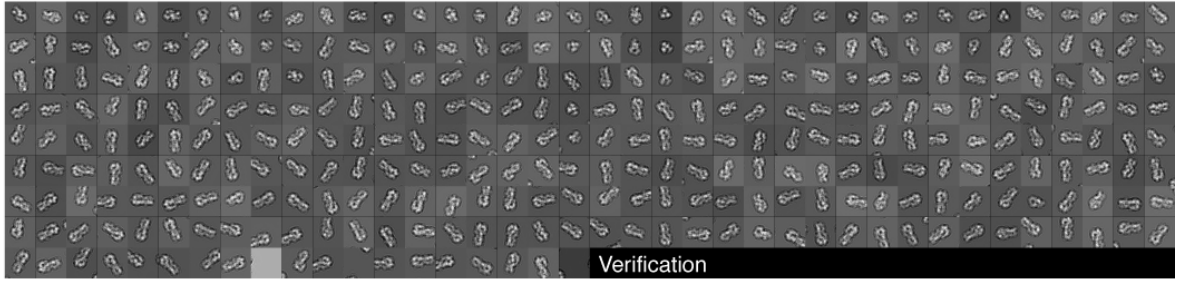
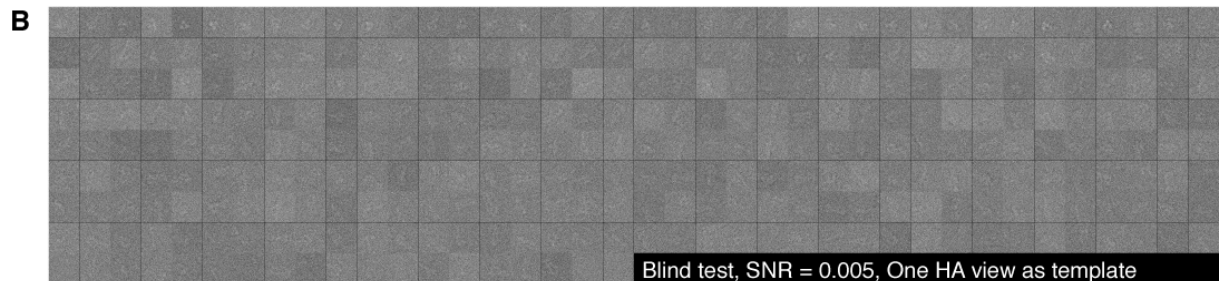
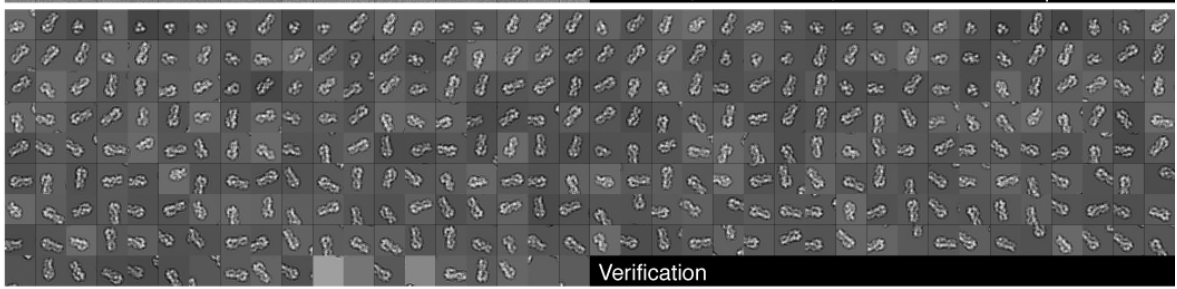
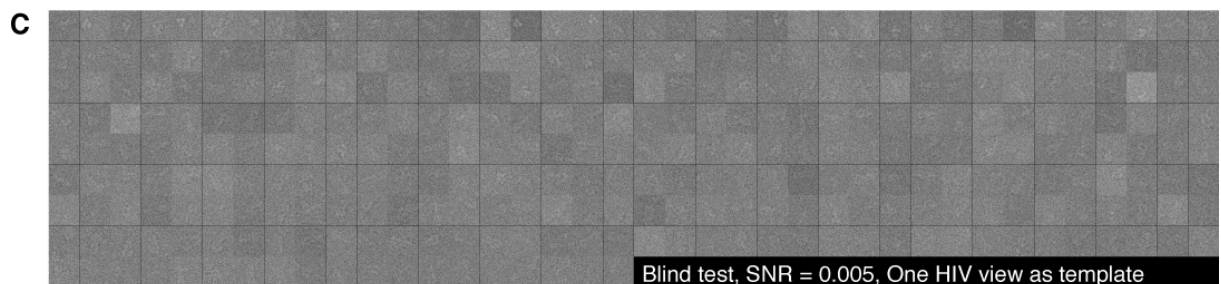
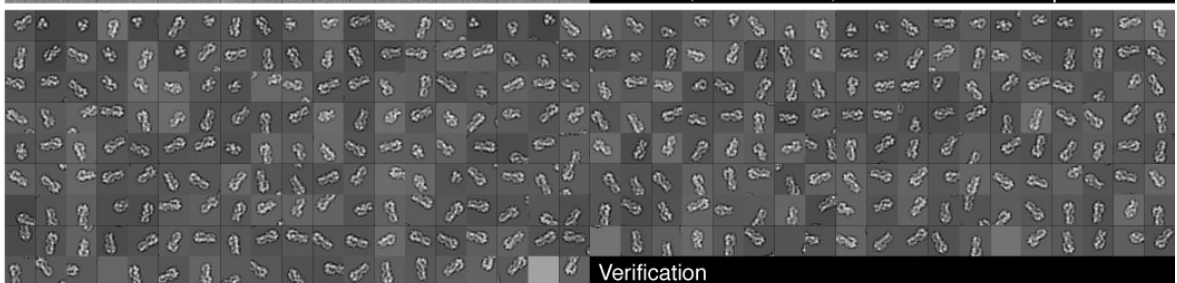

**Figure 6**. Comparison of the FLC-based particle-picking results near the critical SNR with different templates. In the upper half of each panel, a gallery of 323 noisy particles boxed out of the influenza virus HA-containing micrographs with an SNR of 0.005 are shown. The lower half of each panel shows a gallery of noiseless particles picked out of the original noiseless micrograph, using the same boxing parameters and in the same sequence as in the corresponding upper panel. This comparison provides a visual verification of the particle-picking performance. The particle-picking templates were a Gaussian circle (A), one projection view of the influenza virus HA trimer (B) and one projection view of the HIV-1 envelope glycoprotein trimer (C).

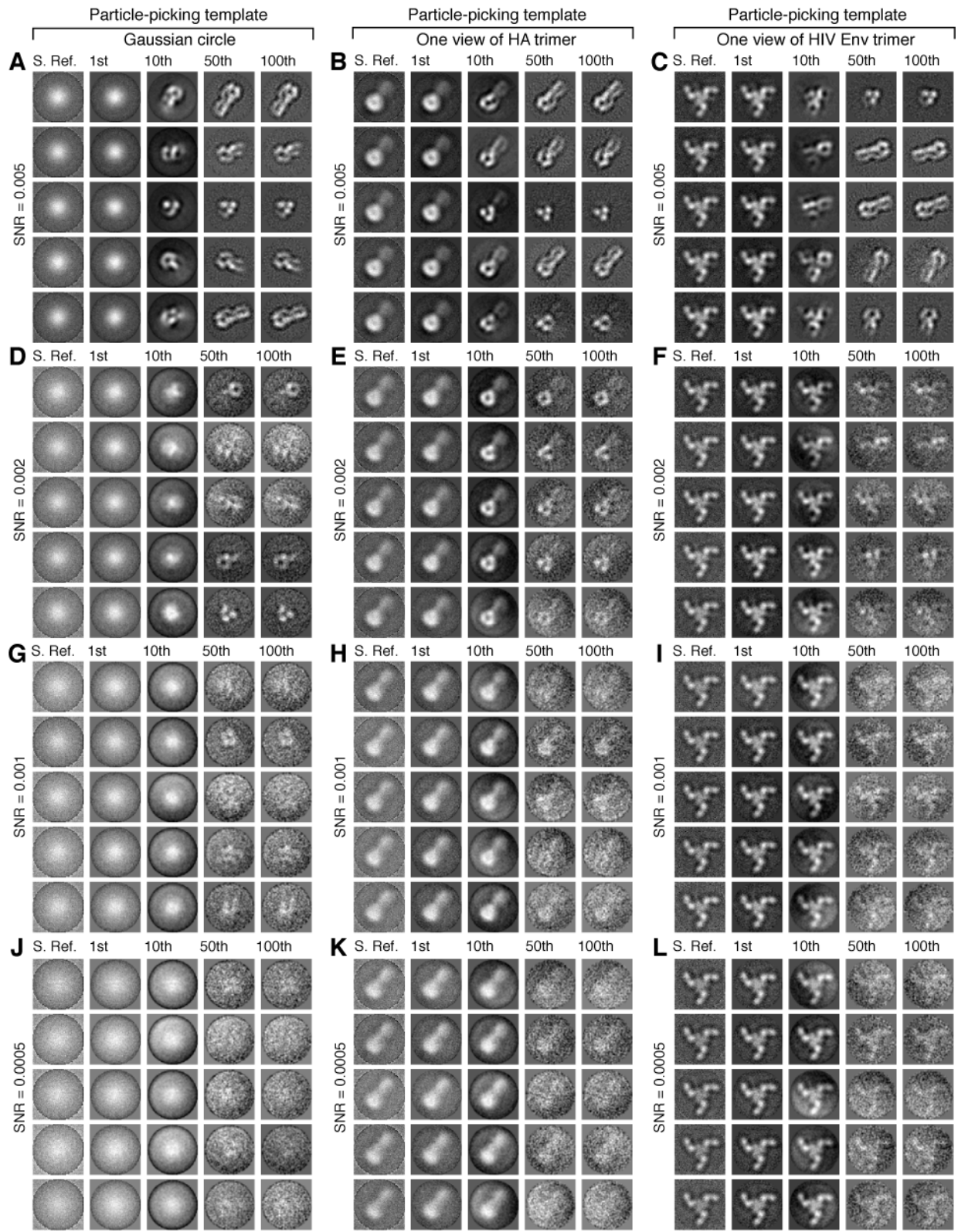

**Figure 7**. Effects of the particle-picking template used in FLC and the micrograph SNR on ML optimization. Noisy micrographs containing the influenza virus HA trimers with different SNRs were subjected to DTF testing, using different templates for particle picking. The corresponding SNRs of the micrographs from which the particle sets were picked are 0.005 (A, B and C), 0.002 (D, E and F), 0.001 (G, H and I) and 0.0005 (J, K and L). The templates used in particle picking were a Gaussian circle (A, D, G and J), one projection view of the influenza virus HA trimer (B, E, H and K) and one projection view of the HIV-1 envelope glycoprotein trimer (C, F, I and L). The particles picked by FLC were randomly divided into five classes and averaged; these "class averages" are shown in the leftmost column of each panel A-L. Using the random class averages as starting references, each assembly of data sets was subjected to multi-reference ML classification. In each panel, the five rows of image series correspond to five particle orientation classes generated by ML, with the starting reference (S. Ref) and class averages of the milestone iterations (1st, 10th, 50th, and 100th) shown in a row. The DTF testing results show that ML optimization can recover the weak signal of the influenza virus HA trimer if there is sufficient SNR in the images.

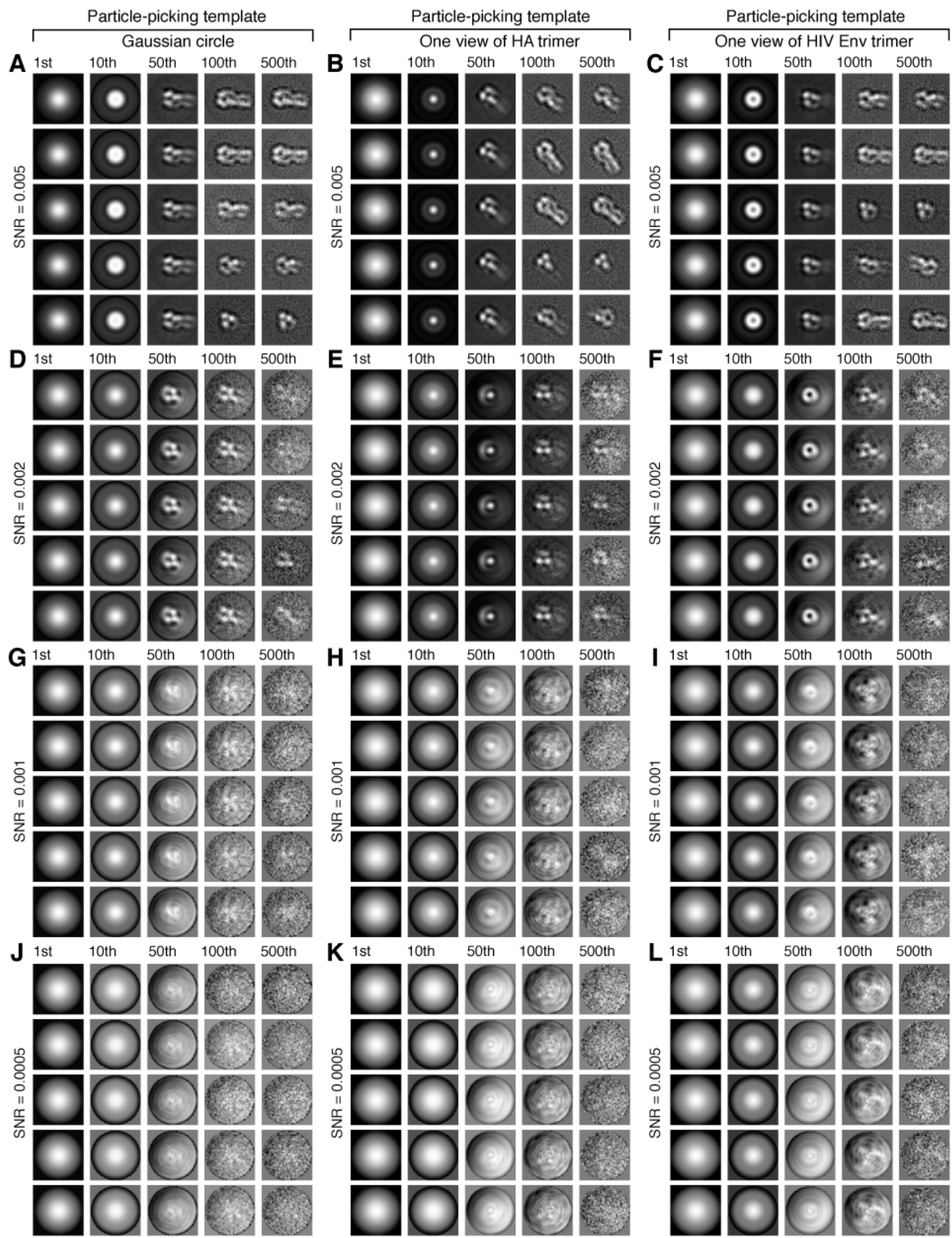

**Figure 8**. Effects of using a Gaussian circle as a starting reference for ML optimization. The procedures shown in Fig. 7 were repeated with a Gaussian circle as the starting reference for all the data sets in the multi-reference ML classification. The corresponding SNRs of the micrographs from which the particle sets were picked are 0.005 (A, B and C), 0.002 (D, E and F), 0.001 (G, H and I) and 0.0005 (J, K and L). The templates used in particle picking were a Gaussian circle (A, D, G and J), one projection view of the influenza virus HA trimer (B, E, H and K) and one projection view of the HIV-1 envelope glycoprotein trimer (C, F, I and L). In each panel, the five rows of image series correspond to five particle orientation classes generated by ML, with the class averages of the milestone iterations ($1^{st}$, $10^{th}$, $50^{th}$, $100^{th}$, $500^{th}$) shown in a row. The results show that the particle-picking template is not recapitulated by the ML optimization when a Gaussian circle is used as a starting reference.

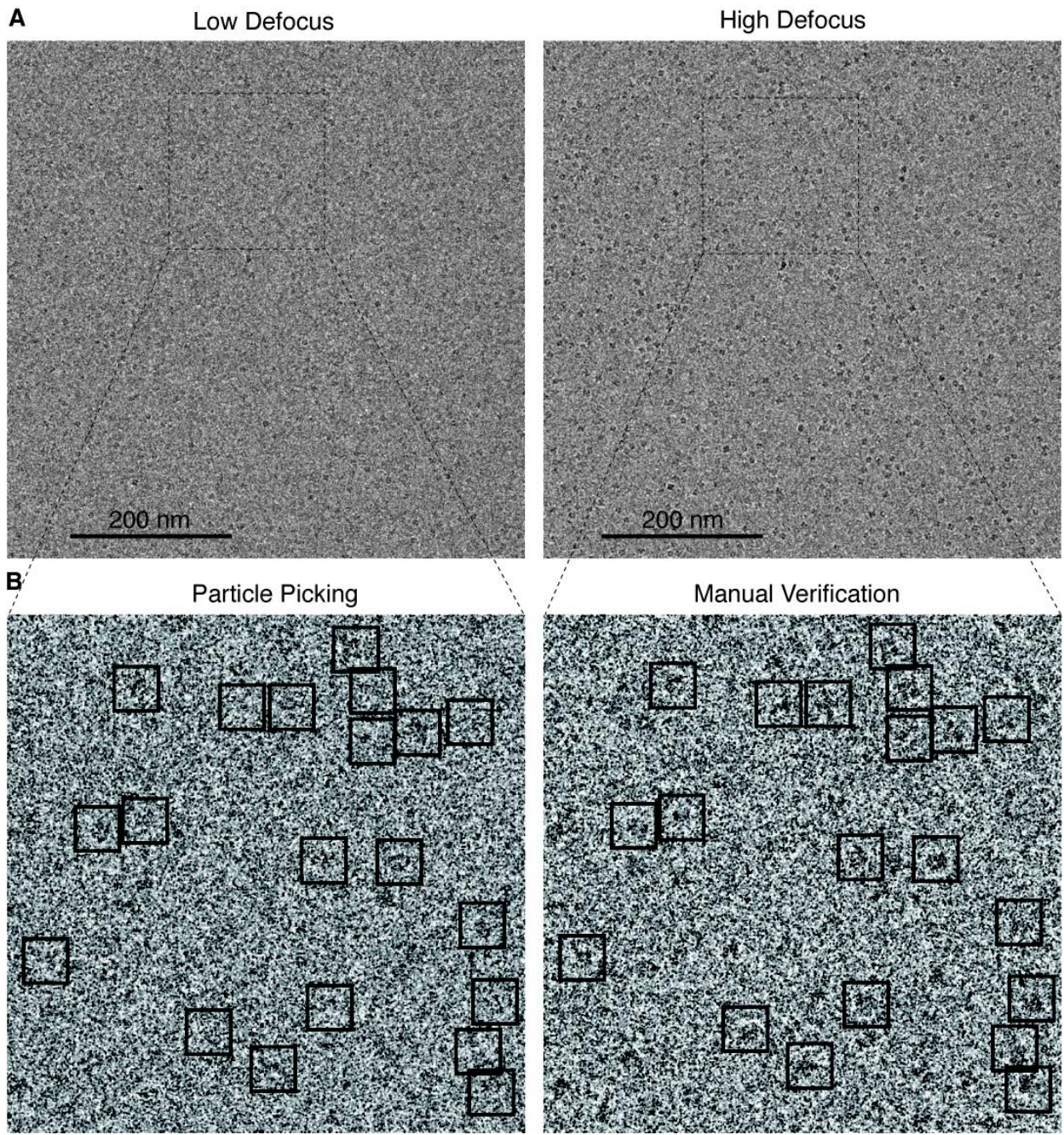

**Figure 9**. Automated particle picking from low-defocus (close-to-focus) micrographs and manual verification of picked particles from high-defocus (far-from-focus) micrographs. (A) Typical focal pair of micrographs of the 173-kDa glucose isomerase complex taken on a Gatan K2 Summit direct detector camera in the electron counting mode at a defocus of -1.8 μm (left) and -3.3 μm (right). The dimensions of the micrograph are 3696 x 3696 pixels, with a pixel size

of 1.74 Å. Each micrograph was exposed to a dose of 10 electrons / Å$^2$. (B) An expanded view of the particle-picking boxes mapped on the low-defocus micrograph on the left, after automated particle picking by the FLC algorithm was applied to this low-defocus micrograph. In this case, a Gaussian circle was used as the particle-picking template. The same set of boxes is mapped to the higher-defocus micrograph taken in the same sample area for manual verification. The particles picked from the low-defocus micrograph (left) are of low contrast. Nonetheless, the picked particles are mostly true particles, as manually verified by the high-defocus micrograph (right). Note that the particle position moved ~8 nm in the high-defocus micrograph compared to the low-defocus micrograph due to a minor imperfection in the alignment of the rotation center and the camera center. This small movement of the images makes it difficult to directly use the coordinates of the boxed particles from the high-defocus micrograph to pick particles from the low-defocus micrograph without additional alignment of the focal-pair micrographs.

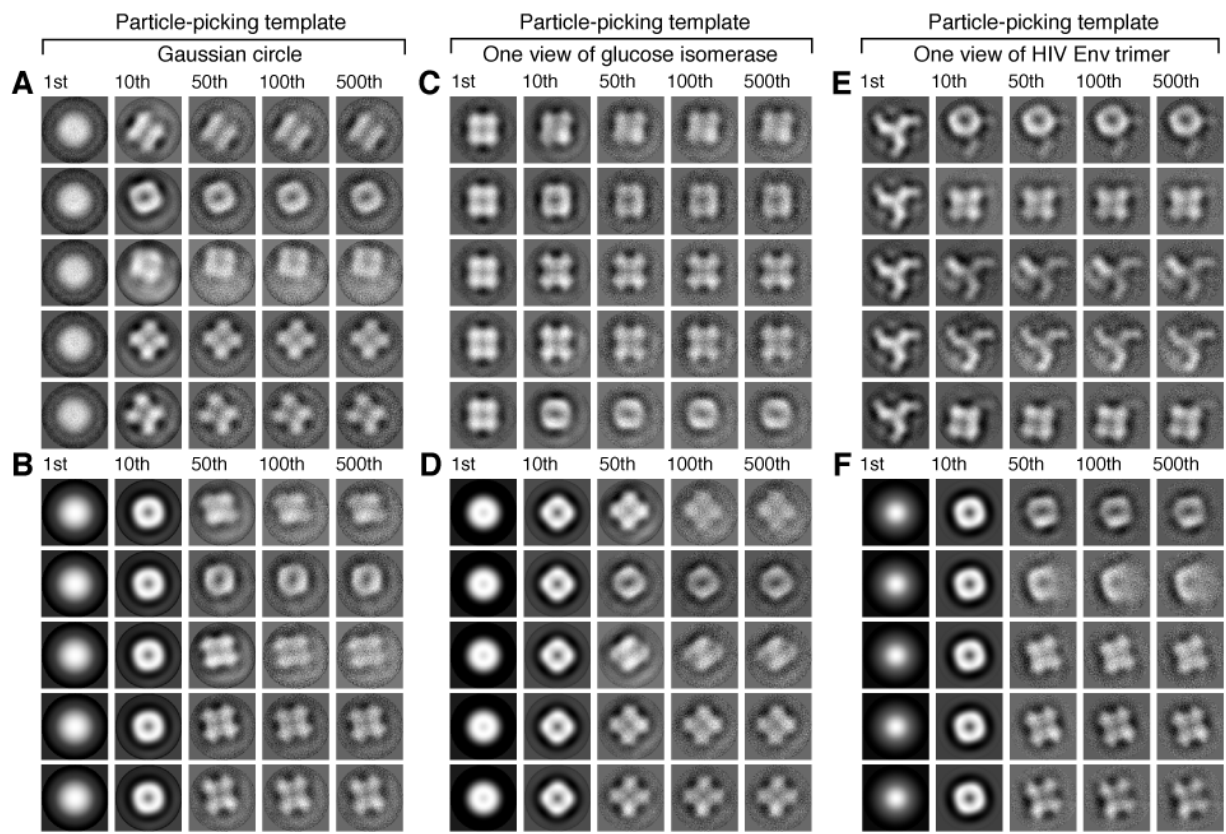

**Figure 10**. Effects of different particle-picking templates and starting references in ML optimization of real cryo-EM images of the glucose isomerase complex. The templates used in particle picking were a Gaussian circle (A, B), one projection view of the glucose isomerase complex (C, D) and one projection view of the HIV-1 envelope glycoprotein trimer (E, F). The approximate percentage of false-positive particles assembled in the three cases, estimated through the manual examination of the larger-defocus micrographs in the focal pairs, was 6% (A, B), 4% (C, D) and 11% (E, F). In the ML optimization step, the unaligned averages of randomly classified particles were used as starting references in panels A, C and E, and a Gaussian circle was used as the starting reference in panels B, D and F. In each panel, the five rows of image series correspond to five particle orientation classes generated by ML, with the class averages of the milestone iterations ($1^{st}$, $10^{th}$, $50^{th}$, $100^{th}$, $500^{th}$) shown in a row.

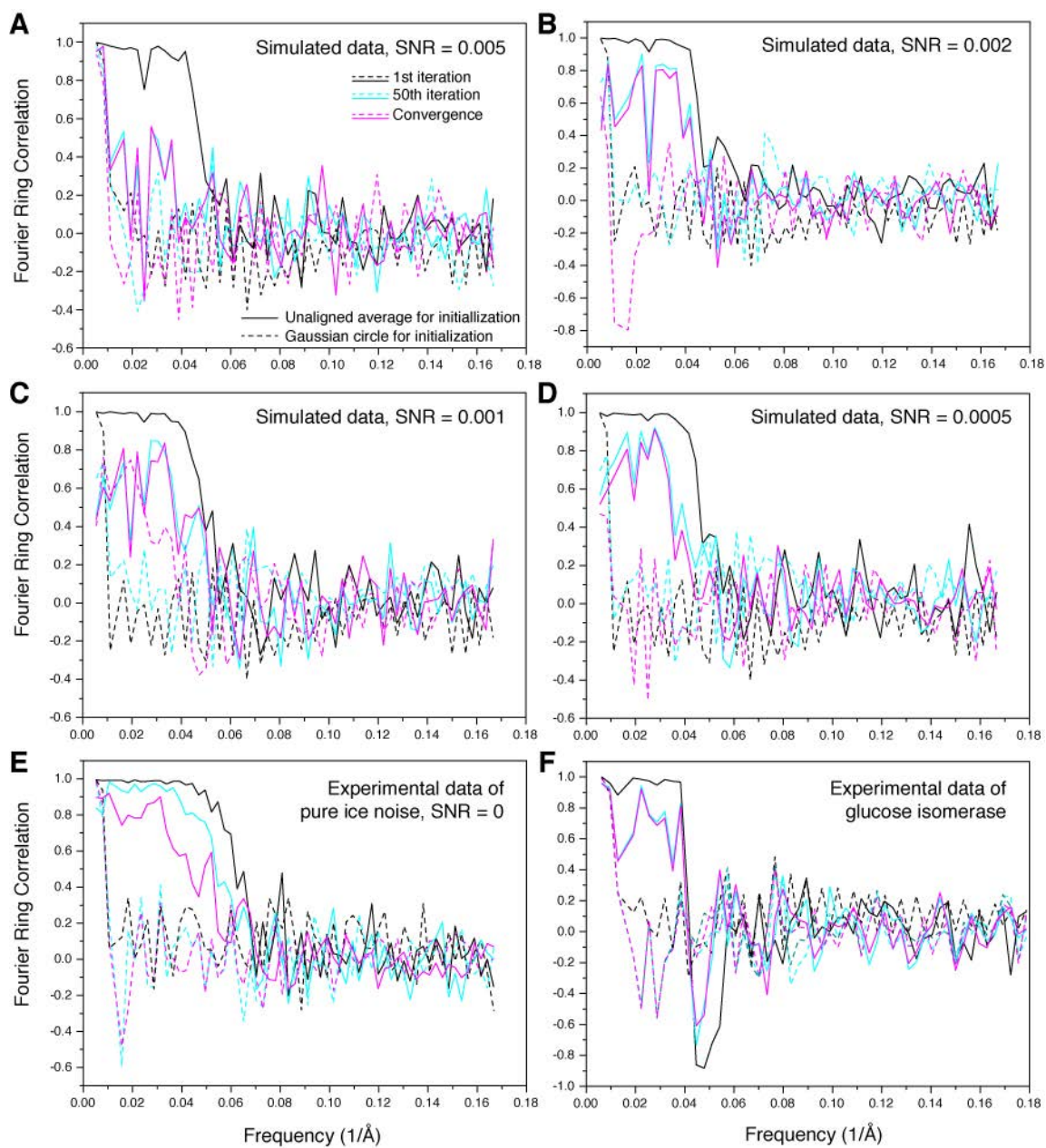

**Figure 11**. Fourier ring correlation (FRC) between the class averages and the particle-picking template of the HIV-1 Env trimer. Panels A-D show the results using the class averages of the simulated data of the HA trimer, with SNR = 0.005 (A), 0.002 (B), 0.001 (C) and 0.0005 (D), corresponding to the results shown in Fig. 7C, F, I and L and Fig. 8C, F, I and L. (E) shows the results using the pure ice noise data in the absence of any proteins, as demonstrated in Fig. 2F

and G. (F) shows the results using the real cryo-EM data of the 173-kDa glucose isomerase complex, corresponding to Fig. 10E and F. The solid and dashed curves were computed from the class averages from ML optimization using the unaligned averages and a Gaussian circle as starting references, respectively. The color indicates the iteration of ML optimization at which the class average was computed. For each case, the FRC analysis is shown for a single class average; the results were similar for other class averages in each case.